\documentclass[a4paper,10pt]{edpsci} 

\usepackage{graphicx}
\usepackage{epstopdf}
\usepackage[numbers]{natbib}
\usepackage{hyperref}
\usepackage{url}
\usepackage{CJKutf8}
\usepackage{arabtex}

\usepackage[T2A]{fontenc}
\usepackage[utf8]{inputenc}
\usepackage{amsmath,graphicx,amsfonts,bm, mathtools}
\usepackage{amssymb}
\usepackage[normalem]{ulem} 
\usepackage{tikzsymbols}
\usepackage{tabularx}
\usepackage{pgfplots}
\usepackage{url}
\usepackage{tikz}
\usepackage{ulem}
\usepackage{makecell}
\usepackage{subcaption}
\usepackage[T3,T1]{fontenc}
\usepackage{xfrac}
\usetikzlibrary{pgfplots.groupplots} 
\usetikzlibrary{mindmap,shadows}

\usepackage{siunitx} \usepackage{xcolor,cancel}
\usepackage[most]{tcolorbox}

\newtcbtheorem[]{mylemmma}{Lemma}{colframe=green,colback=white, width=\textwidth, left=0pt}{lem}

\tcbset{colback=yellow!10!white, colframe=red!50!black, 
	highlight math style= {enhanced, 
		colframe=red,colback=red!10!white,boxsep=0pt}
}

\newcommand{\bioss}{BIOSS}

\newcommand{\rrec}{\ensuremath{\mathbf{r}}}
\newcommand{\rb}{\ensuremath{\mathbf{b}}}
\newcommand{\rbeta}{\ensuremath{\boldsymbol{\beta}}}

\newcommand{\rsrc}{\ensuremath{\mathbf{u}}}

\newcommand{\pscat}{\ensuremath{{p_{\text{{scat}}}}}}
\newcommand{\pinc}{\ensuremath{{p_{\text{inc}}}}}

\newcommand{\domain}{\ensuremath{\Omega}}
\newcommand{\cdom}{\ensuremath{\Omega_{c}}}
\newcommand{\bnd}{\ensuremath{\Gamma}}

\newcommand{\Ros}{{R}_{\rb \rsrc}}
\newcommand{\Ror}{{R}_{\rrec \rsrc}}
\newcommand{\Rsr}{{R}_{\rrec \rbeta}}
\newcommand{\Rbb}{{R}_{\rb \rbeta} }

\newcommand{\inputsrc}{\ensuremath{{x}}}
\newcommand{\pbndry}{\ensuremath{{q}}}
\newcommand{\weightbie}{\ensuremath{{w}}}

\newcommand{\nvct}{\mathbf{n}} 

\newcommand{\thetabetar}{{\theta}_{\rrec \rbeta }}
\newcommand{\thetabb}{{\theta}_{\rb \rbeta } }

\newcommand{\gbr}{{g}_{\rrec \rbeta }}
\newcommand{\hbr}{{h}_{\rrec \rbeta }}

\newcommand{\gbb}{{g}_{\rb \rbeta }}
\newcommand{\hbb}{{h}_{\rb \rbeta }}

\newcommand{\Asstd}{\mathcal{A}} 
\newcommand{\Bsstd}{\mathcal{B}} 
\newcommand{\Csstd}{\mathcal{C}} 
\newcommand{\Dsstd}{\mathcal{D}} 
\newcommand{\Isstd}{\mathcal{I}} 

\newcommand{\Assp}{\mathbf{A}} 
\newcommand{\Bssp}{\mathbf{B}} 
\newcommand{\Cssp}{\mathbf{C}} 
\newcommand{\Dssp}{\mathbf{D}} 
\newcommand{\Issp}{\mathbf{I}} 

\newcommand{\q}{\mathbf{q}}

\definecolor{mycolor1}{RGB}{178,171,210}
\definecolor{mycolor2}{RGB}{253,184,99}
\definecolor{mycolor3}{RGB}{230,97,1}
\definecolor{mycolor4}{RGB}{94,60,153}
\definecolor{mycolor5}{RGB}{166,219,160}
\definecolor{mycolor6}{RGB}{0,136,55}

\hypersetup{colorlinks=true,citecolor=blue,urlcolor=blue,linkcolor=blue}

\usepackage[ruled]{algorithm2e}

\makeatletter
\protected\def\begin#1{%
  \UseHook{env/#1/before}%
  \@ifundefined{#1}%
    {\def\reserved@a{\@latex@error{Environment #1 undefined}\@eha}}%
    {\def\reserved@a{\def\@currenvir{#1}%
        \edef\@currenvline{\on@line}%
        \@execute@begin@hook{#1}%
        \csname #1\endcsname}}%
  \@ignorefalse
  \begingroup
  \let\end\a@l@end 
  \@endpefalse\reserved@a}
\makeatother

\begin{document}

\title{A state-space representation of the boundary integral equation for room acoustic modelling}




\author{Randall Ali\inst{1}\correspondingauthor{\email{r.ali@surrey.ac.uk}}
\and
Thomas Dietzen\inst{2}
\and
Matteo Scerbo\inst{1}
\and  Enzo De Sena\inst{1}
\and Toon van Waterschoot\inst{2}  }




\institute{Institute of Sound Recording, University of Surrey, Guildford, UK\and
STADIUS, Dept. of Electrical Engineering, KU Leuven, Leuven, Belgium}

\abstract{We introduce a new framework for room acoustics modelling based on a state-space model of the boundary integral equation representing the sound field in a room. Whereas state-space models of linear time-invariant systems are traditionally constructed by means of a state vector and a 4-tuple of system matrices, the state-space representation introduced in this work consists of a state function representing the pressure distribution at the room boundary, and a 4-tuple of integral operators. We refer to this representation as a boundary integral operator state-space (BIOSS) model and provide a physical interpretation for each of the integral operators. As many mathematical operations on vectors and matrices translate to functions and operators, the BIOSS representation can be manipulated to obtain two transfer function representations, having either a feedback or a parallel feedforward structure. Consequently, various equivalent representations for room acoustics are obtained in the BIOSS framework, in the time or frequency domain, and in continuous or discrete space. We discuss two future directions for how the proposed framework can be fertile for research on room acoustics modelling. Firstly, we identify equivalences between the BIOSS framework and various existing room acoustics models (boundary element models, delay networks, geometric models), which may be used to establish relations between existing models and to develop novel room acoustics models. Secondly, we postulate on how concepts from state-space theory, such as observability, controllability, and state realization, can be used for developing new inference and control methods for room acoustics.}


\maketitle

\section{Introduction}

Room acoustic modelling, i.e., the modelling of acoustic wave fields within bounded interior domains (e.g., concert halls, living rooms, and classrooms) has significantly contributed to our understanding of complex acoustic behaviour within enclosed spaces. With the ability to simulate sounds at receiver positions within a room \cite{Vorlander_2013}, estimate source signals from room acoustic measurements (an inverse problem) \cite{Antonello19}, or even estimate room acoustic parameters (e.g., room geometry, absorption coefficients of materials, etc.) \cite{Dokmanic2013}, room acoustic models have been central to a wide range of applications. These include building design \cite{long2005architectural}, auralization \cite{vorlander2020auralization}, sound design for video games \cite{raghuvanshi2007real}, music production \cite{Valimaki50yrs}, virtual/augmented reality \cite{kim2019immersive}, among many more. Given the diversity of applications and requirements of various users, however, there is no single all-encompassing room acoustic model that the community can adopt. Rather, decades of research has led to a number of different frameworks, i.e., a set of assumptions and constraints, from which practical room acoustic models are developed, often with tradeoffs between physical accuracy and computational complexity tailored to the application at hand. While we do not expect that a universally-applicable room acoustic model can be developed, what seems to be missing from our current slate of room acoustic models are connections and relationships among them, as well as one that has flexibility to incorporate varying degrees of physics principles and room acoustic measurement data, particularly (spatial) room impulse responses ((S)RIRs). 

Frameworks for developing room acoustic models can be categorized along several dimensions. For instance, one dimension can represent the degree to which models are based on physics-first principles and approximations thereof (physical models) or on room acoustic measurements (data-driven models). A second dimension can indicate whether a model assumes a wave-based \cite{hamilton2016finite} or geometric acoustic propagation \cite{Savioja2015}, and a third can correspond to the specific way in which room acoustic models are implemented, i.e., using input-output relations (e.g., transfer functions) or a state-space formulation \cite{willems1997introduction}. Room acoustic models may be developed using frameworks at various intersections of these dimensions, so for instance, one may develop a room acoustic model that is physics-based, adheres to geometric acoustic propagation, and is implemented as a state-space model \cite{scerbo2024modartmodaldecompositionacoustic}. As previously mentioned, a choice of framework is usually determined by the requirements of the application. A system that needs to run in real-time may therefore favour geometric acoustic propagation over wave-based propagation due to its reduced computational complexity, but at the same time will have to work within the limits of the physical accuracy of such a model.

In this paper, we are presenting a framework for developing room acoustic models which stems from a physical model, but allows for the incorporation of data, is inherently wave-based, and uses a state-space formulation. The advantage of this framework is that it creates a path towards making connections among various room acoustic models which use alternative frameworks, and it can be adapted to be solely physics-based, data-driven, or somewhere in between. In order to further contextualize the contribution and how it fits within the orbit of existing room acoustic models, we will firstly elaborate on the aforementioned categorized frameworks.  

Many physical models are predicated on continuous space-time mathematical descriptions of the underlying physics such as the Kirchhoff Helmholtz integral equation or the 3D partial differential wave equation \cite{jacobsen2013, Pierce81}. Discretization of these equations in space and time yield high-dimensional, structured linear systems of equations which can then be numerically solved leading to the boundary element method, the finite element method, or the finite difference time domain method \cite{bergman2018computational}. The dimensions of the matrices for solving these linear systems of equations can, however, become quite large as the resolution of the spatial discretization increases, restricting this wave-based approach to computationally viable cases, where, for instance, only lower and/or individual frequencies are considered. To relieve this computational constraint, physical models may be alternatively based on geometric acoustic propagation \cite{Savioja2015} where the dimensions of the room and its walls are large compared to the wavelength of the sound \cite{kuttruff2009room}, such as the image source method \cite{Allen1979}, ray tracing \cite{krokstad1968calculating}, beam tracing \cite{funkhouser2004beam}, and acoustic radiance transfer \cite{siltanen2007room}. Attempts have also been made at hybrid approaches between wave-based and geometric frameworks in order to have the advantages of both \cite{aretz2009combined, Wittebol2024}, and to investigate appropriate transition criteria between the two frameworks \cite{ali2023relating}. Whether wave-based, geometric, or hybrid, one commonality in the physics-first approach is that it facilitates the rendering of the acoustics within a room for arbitrary source and receiver positions using a single model, whose parameters are related to the physical properties of the acoustic scenario (e.g., absorption coefficients, bidirectional reflectance distribution functions \cite{Savioja2015}, etc.). Hence a compact room acoustics model can be attained, but it comes at the cost of prior knowledge of these model parameters, which is not always available \cite{jeong2022design}.

Data-driven models are often based on (S)RIRs \cite{kuttruff2009room, merimaa2005spatial}. RIRs are rich in information, as they represent a superposition of room-specific acoustic events, which to some extent are separable in space and time. They can also be measured using well-established procedures \cite{farina2007advancements} and are favoured among signal processing practitioners as they can be implemented as digital finite impulse response filters \cite{Oppenheim2009}. Discretization is also not an issue as temporal discretization is handled by Nyquist-Shannon sampling theory commonly used in digital signal processing \cite{Oppenheim2009}, and spatial discretization is implicit by the point-to-point nature of RIRs, considering sources and receivers as point observers. Other data-driven models include modal response models, e.g. common-acoustical-pole and zero \cite{haneda1994common}, orthonormal basis function \cite{vairetti2017scalable}, and parallel filter models \cite{ramo2014high}. Unlike physical models, data-driven models do not need to assume any specific prior knowledge as measurements are sufficient to identify RIRs between predetermined source and receiver positions. A complete room characterization, however, requires the measurement and identification of RIRs between many source and receiver pairs, which is inefficient from a measurement and modelling point of view. Some data-driven models, however, have been shown to be more effective for room acoustics applications when a physical prior is included in the model structure \cite{van2008optimally}. A geometric prior is used in RIR decomposition models such as the spatial decomposition method \cite{tervo2013spatial}, whereas wave-based priors are used in wave decomposition models, e.g. wave field analysis \cite{berkhout1997array}, plane-wave decomposition \cite{pinto2010space}, and spherical-harmonic decomposition \cite{jarrett2017theory}.

The resulting room acoustic models from some of these approaches can be implemented in an input-output manner, i.e., an input is specified (e.g., some human speech) and an output (the sound at the receiver) is generated without requiring knowledge of any latent variables within the system. For example, given a RIR, a simple convolution with an input signal can be done to yield an output at some position within the room. The other approach to implementation is via a state-space model \cite{willems1997introduction}, which can be used to represent any linear time-invariant (LTI) system. This type of implementation can be computationally efficient, provides further insight into the mechanisms responsible for generating a particular acoustic field if physically motivated \cite{scerbo2024modartmodaldecompositionacoustic}, and different state-space models can even result in the same input-output behavior. Room acoustic models based on delay networks \cite{Valimaki50yrs}, such as the feedback delay network (FDN) \cite{jot1991digital} and scattering delay network (SDN) \cite{DeSena2015} are typically implemented using state-space formulations. Such models employ a geometric acoustics framework \cite{scerbo2024common} and can be designed from physics-first principles \cite{DeSena2015}. State-space models can also incorporate (S)RIR data, which can subsequently be used to estimate the state-space parameters of an FDN \cite{Mezza2024, santo2024feedback}. Alternatively, RIR data can be used in a state-space framework to build models for moving sources/receivers \cite{macwilliam2024state} or even for other types of room acoustic applications such as echo cancellation \cite{enzner2010bayesian}.

A state-space framework can therefore be useful in addressing the apparent gap between physical models where prior knowledge of room acoustic parameters is required and data-driven models where only a finite set of acoustic measurements ((S)RIRs) is available. Following from this premise, our intention is to leverage a state-space framework as the basis for a room acoustics modelling framework that is capable of bridging physical and data-driven models by efficiently representing the spatiotemporal acoustic behaviour of a room based on a concise set of measurements with minimal prior knowledge. Although this is achieved to some extent already by delay network-based room acoustic models, these models are typically developed with a geometrical acoustics lens, from which further physics principles can be subsequently incorporated \cite{DeSena2015}. In this work, we alternatively start from a physical model, from which state-space and wave-based perspectives are imposed. 

More specifically, we propose a room acoustics framework that is based on the interpretation of the time-domain Kirchhoff Helmholtz integral equation or boundary integral equation (BIE) \cite{Pierce81, wu2000boundary, Hargreaves2007} as a state-space model. We initially formulate this in continuous-time and continuous-space for a single-input single-output LTI using what we refer to as an \emph{operator state-space model}, which makes use of integral operators. We will subsequently refer to this as the boundary integral operator state-space model (\bioss). By assuming a locally reacting surface, we interpret the acoustic pressure on the boundary of an interior domain as the state of the system. Using the time-domain BIE, a state equation is then derived where the initial state is updated via a state-transition integral operator, which defines how the state is scattered among boundaries within the room. The time-domain BIE is then used once again to derive a measurement/observation equation which is a sum of the incident acoustic pressure component from the source and a scattered acoustic pressure component, the latter of which uses another integral operator to define how the state propagates from the boundaries to the receiver.

Starting from the \bioss, the immediate thought is to discretize in space and time in order to obtain a room acoustic model that can be implemented in practice. As would be expected, this will effectively lead to the time-domain boundary element method (TD-BEM) \cite{Hargreaves2007, wu2000boundary}, which is known to exhibit stability issues \cite{Rynne1985, Smith01101990}. There have, however, been several propositions for circumventing these instabilities using various time discretization strategies \cite{Sadigh1992, ergin2002plane}, including that of convolution quadrature \cite{lubich1994multistep, banjai2014fast, sauter2017convolution}. Nevertheless, what is important to realize here, and the hypothesis we put forth in this paper, is that the \bioss{} framework is more than just a starting point for the TD-BEM. Rather it is a framework that can be used to potentially address TD-BEM instabilities, derive new room acoustic models, find connections to existing models, and incorporate data to estimate model parameters.

This paper is organized as follows. In Section 2, we review the fundamentals of the time-domain BIE for the interior scattering problem. In section 3, we develop the continuous space \bioss{} framework both in the continuous time domain and the Laplace domain. In section 4, we describe the spatial discretization of the \bioss{} yielding an integral-operator-free state-space model. In section 5, we propose various research problems for which the framework could be useful. We conclude the paper in section 6.

\section{The boundary integral equation}

\begin{figure}
	\centering
	\includegraphics[width=\linewidth]{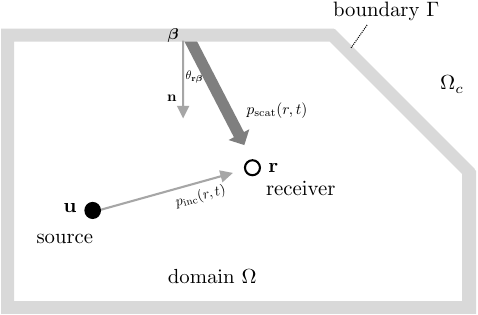}
	\caption{Conceptual illustration of the acoustic scenario considered for wave propagation within the interior domain, $\domain$, enclosed by the boundary, $\bnd$. The pressure field at the receiver position, $\rrec$, and time, $t$, consists of a scattered pressure contribution, $\pscat(\rrec, t)$ due to boundary interactions, and an incident pressure contribution, $\pinc(\rrec, t)$ from a source located at $\rsrc$. $\Omega_c$ is the complementary domain, $\nvct$ is a normal vector pointing into $\domain$, and $\rbeta$ represents the position on the boundary.}
	\label{fig:room_scene}
\end{figure}

We consider a domain, $\domain$, enclosed by the boundary $\bnd$,  and a complementary domain, $\cdom$ as depicted in Fig. \ref{fig:room_scene}, where our objective is to obtain the pressure field at some position, $\rrec$, and time, $t$, within $\domain \setminus \bnd$ due to a source excitation also within $\domain \setminus \bnd$. As depicted in Fig. \ref{fig:room_scene}, the pressure field at $\rrec$ can be conceptualized as consisting of two components: (i) the scattered wave field, $\pscat$, resulting from acoustic interactions with the boundary, and (ii) the incident wave field (direct component), $\pinc$ from a source within $\domain \setminus \bnd$.

The scattered wave field can be obtained using the Kirchhoff Helmholtz integral equation \cite{wu2000boundary, jacobsen2013, Pierce81}, or what we will simply refer to as the boundary integral equation (BIE). In the time-domain, for a normal vector pointing into $\domain$, the BIE is 
\begin{align}
	\label{eq:KHIT_TD}
	w(\rrec)\pscat(\rrec,t) =& \int_{0}^{t} \int_{\bnd}  \pbndry(\rbeta, \tau)   \frac{\partial G}{\partial \nvct }(\rrec, t | \rbeta, \tau)  d\bnd \hspace{0.1cm} d\tau \nonumber \\ &- \int_{0}^{t} \int_{\bnd} \frac{\partial \pbndry}{\partial \nvct}(\rbeta, \tau) G(\rrec, t | \rbeta, \tau) \hspace{0.1cm} d\bnd \hspace{0.1cm} d\tau, 
\end{align}
where $\rbeta$ is a position on the boundary $\bnd$, $\pbndry(\rbeta, \tau)$ is an acoustic pressure ($\si{Pa}$) on the boundary,  $\partial \pbndry (\rbeta, \tau) / \partial \nvct$ is its normal derivative, and for smooth boundaries\footnote{For non-smooth boundaries such as corners, the value for $\weightbie(\rrec)$ can be obtained by considering the integral of the normal derivative of the fundamental solution of the Laplace equation \cite{wu2000boundary}.}, $\weightbie(\rrec)$ is defined as
\begin{align}
	\label{eq:KHIT_scaling}
	\weightbie(\rrec)  &=  \begin{cases}
		4 \pi \hspace{0.5cm} \text{if $\rrec \in \domain \setminus \bnd$} \\
		2 \pi  \hspace{0.5cm} \text{if $\rrec \in \bnd$} \\
		0 \hspace{0.7cm} \text{if $\rrec \in \cdom$}.
	\end{cases}
\end{align}

\noindent $G(\rrec, t | \rbeta, \tau) $ is a spatio-temporal Green's function, where the notation relates propagation from the spatio-temporal coordinates $(\rbeta, \tau)$ towards $(\rrec, t)$. $G(\rrec, t | \rbeta, \tau)$, also referred to as the time-domain Green's function, is the fundamental solution to the inhomogenous wave equation
\begin{align}
	\label{eq:inhomhelmgreentd}
	\nabla^{2}G(\rrec, t | \rbeta, \tau)  - \frac{1}{c^{2}}\frac{\partial^{2}}{\partial t^{2}} G(\rrec, t | \rbeta, \tau)   &= -4 \pi \delta(\rrec - \rbeta) \delta(t - \tau), 
\end{align}
where $c$ is the speed of sound in air (m/s) and $\delta(\rrec - \rbeta) \delta(t - \tau)$ represents an impulse in both space and time. For $\rrec \in \mathbb{R}^{3}$, it is given by
\begin{align}
	\label{eq:GreenfcnTD}
	G(\rrec, t | \rbeta, \tau) = \frac{1}{\Rsr} \delta\left(t - \frac{\Rsr}{c} - \tau \right)
\end{align}
where $\Rsr$ is a shorthand notation for $\Rsr=~R(\rrec, \beta)~=~\|\rrec - \rbeta\|_{2}$. The  normal derivative of $	G(\rrec, t | \rbeta, \tau)$ is also given by \cite [Ch. 8]{wu2000boundary}
\begin{align}
	\label{eq:dGdn_td}
	\frac{\partial G }{\partial \nvct} (\rrec, t | \rbeta, \tau) =& - \frac{\partial \Rsr }{\partial \nvct}  \frac{1}{c\Rsr}  \frac{\partial \delta}{\partial t} \left(t - \frac{\Rsr}{c}  - \tau \right)   \nonumber \\
	&- \frac{\partial \Rsr }{\partial \nvct}   \frac{1}{\Rsr^{2}}\delta\left(t - \frac{\Rsr}{c} - \tau  \right)  ,
\end{align}
where ${\partial \Rsr }/{\partial \nvct}$ is the directional derivative in the normal direction
\begin{align}
	\label{eq:drdn_td}
	\frac{\partial \Rsr }{\partial \nvct} = \frac{(\rrec -\rbeta)^{T}  \mathbf{n} } {\Rsr } = \cos \thetabetar
\end{align}
where $ \mathbf{n}$ is the unit normal vector and $\thetabetar$ is a shorthand notation for $\theta(\rrec, \rbeta)$, the angle between the vectors $(\rrec -\rbeta)$ and $\mathbf{n}$ (depicted in Fig. \ref{fig:room_scene}). Substitution of \eqref{eq:GreenfcnTD}, \eqref{eq:dGdn_td}, and \eqref{eq:drdn_td} into \eqref{eq:KHIT_TD}, results in:
\begin{align}
	\label{eq:KHIT_TD_GFsub}
	&\weightbie(\rrec)\pscat(\rrec,t) = \nonumber\\
	& - \int_{0}^{t} \int_{\bnd}   \pbndry(\rbeta, \tau) \frac{\cos\thetabetar}{c\Rsr}  \frac{\partial \delta}{\partial t}  \left(t - \frac{\Rsr}{c} - \tau \right)    \hspace{0.1cm} d\bnd \hspace{0.1cm} d\tau \nonumber \\
	& - \int_{0}^{t} \int_{\bnd} \pbndry(\rbeta, \tau)  \frac{\cos\thetabetar}{\Rsr^{2}} \delta\left(t - \frac{\Rsr}{c} - \tau \right) d\bnd \hspace{0.1cm} d\tau  \nonumber \\
	& - \int_{0}^{t} \int_{\bnd} \frac{\partial \pbndry (\rbeta, \tau) }{\partial \nvct} \frac{1}{\Rsr} \delta\left(t - \frac{\Rsr}{c} - \tau \right) d\bnd \hspace{0.1cm} d\tau  
\end{align}
where we have deliberately split the scattered acoustic pressure into 3 separate terms.

We can continue to simplify \eqref{eq:KHIT_TD_GFsub} by applying the following sifting and translation properties of the dirac delta function and its derivative for any function, $f$, that is continuously differentiable in time (see equation (3.20) in \cite{Hoskins2011}\footnote{Note that in our case, we have the opposite sign in the integration variable as compared to equation (3.20) in \cite{Hoskins2011}})
\begin{align}
	\label{eq:deltaderiv_trans}
	\int_{-\infty}^{\infty} f(\tau) \delta \biggl(t - \frac{\Rsr}{c} - \tau\biggr) d\tau &=  f\biggl(t - \frac{\Rsr}{c}\biggr), \\
	\label{eq:deltaderiv_trans2}
	\int_{-\infty}^{\infty} f(\tau) \frac{\partial \delta }{ \partial t} \biggl(t - \frac{\Rsr}{c} - \tau\biggr) d\tau &=  \frac{\partial f}{\partial t}\biggl(t - \frac{\Rsr}{c}\biggr).
\end{align}

\noindent Hence \eqref{eq:KHIT_TD_GFsub} reduces to a set of terms involving integrals with respect to space only
\begin{align}
	\label{eq:KHIT_TD_GFsub2}
	\weightbie(\rrec)\pscat(\rrec,t) =& - \int_{\bnd} \frac{\cos\thetabetar}{c\Rsr}  \frac{\partial \pbndry}{ \partial t} \left(\rbeta, t - \frac{\Rsr}{c}\right)      \hspace{0.1cm} d\bnd \hspace{0.1cm}  \nonumber \\
	& -  \int_{\bnd} \frac{\cos\thetabetar}{\Rsr^{2}} \pbndry\left(\rbeta, t - \frac{\Rsr}{c}\right)    d\bnd \hspace{0.1cm}  \nonumber \\
	& -  \int_{\bnd} \frac{1}{\Rsr} \frac{\partial \pbndry }{\partial \nvct}\left(\rbeta, t - \frac{\Rsr}{c}\right)  d\bnd. \hspace{0.1cm}  
\end{align}

\noindent We assume the boundary to be a locally reacting surface, and hence \cite{cox2016acoustic}
\begin{align}
	\label{eq:eulereqn_normal_impdance}
	\frac{\partial \pbndry}{\partial \nvct} (\rbeta, t) = - \frac{\rho}{Z(\rbeta)} \frac{\partial \pbndry}{\partial t} (\rbeta, t),
\end{align}
where $Z(\rbeta)$ is the frequency-independent impedance\footnote{The frequency independency assumption for the impedance is introduced here solely for notational simplicity. More generally, a frequency-dependent boundary condition can be represented by alternatively making use of convolutive operators \cite[Sec. 6.1]{deleforge2025}.} at the boundary relating the pressure and velocity. It is important to note here that the acoustic pressure and velocity at the boundary are not independent quantities but are related by the boundary impedance. Hence, the acoustic pressure and velocity cannot be independently specified on the boundary \cite[Ch.9, sec. 9.6] {jacobsen2013}. In the following, we therefore assume that the boundary impedances are known a priori\footnote{In certain contexts, this is a considerable assumption as it is not trivial to obtain such information. Nevertheless, we keep the assumption for the purpose of developing the framework, from which information regarding boundary impedances can then be reconsidered.}. Substitution of \eqref{eq:eulereqn_normal_impdance} into \eqref{eq:KHIT_TD_GFsub2} then yields
\begin{align}
	\label{eq:KHIT_TD_GFsub_impedance}
	\weightbie(\rrec)\pscat(\rrec,t) =& -\int_{\bnd}  \frac{\cos\thetabetar}{c\Rsr}   \frac{\partial \pbndry}{ \partial t}\left(\rbeta, t - \frac{\Rsr}{c}\right)      \hspace{0.1cm} d\bnd \hspace{0.1cm}  \nonumber \\
	& -  \int_{\bnd} \frac{\cos\thetabetar}{\Rsr^{2}} \pbndry\left(\rbeta, t - \frac{\Rsr}{c}\right)    d\bnd \hspace{0.1cm} \nonumber \\
	& + \int_{\bnd}  \frac{\rho}{\Rsr Z(\rbeta)}  \frac{\partial \pbndry}{ \partial t}\left(\rbeta, t - \frac{\Rsr}{c}\right)      \hspace{0.1cm} d\bnd \hspace{0.1cm} . 
\end{align}
It is worthwhile to highlight that for rigid boundaries, $Z(\rbeta) \rightarrow \infty$, and hence the last term in \eqref{eq:KHIT_TD_GFsub_impedance} will be zero. Finally, the pressure field is then given by
\begin{align}
	\label{eq:total_soln}
	p(\rrec, t) &= \frac{4 \pi}{\weightbie(\rrec)}\pinc(\rrec, t) + \pscat(\rrec, t),
\end{align}
where ${4 \pi}/{\weightbie(\rrec)}$ is a scaling factor and the incident pressure  is 
\begin{align}
	\label{eq:pinc_td1}
	p_{inc}(\rrec,t) &= \int_{0}^{t} G(\rrec, t | \rsrc, \tau)  \inputsrc(\rsrc,\tau) d \tau \nonumber \\
	& =  \frac{1}{\Ror} \int_{0}^{t}  \delta \left(t  - \tau - \frac{ \Ror}{c}\right) \hspace{0.1cm} \inputsrc(\rsrc, \tau)  d \tau \nonumber \\
	& = \frac{1}{ \Ror} \hspace{0.1cm} \inputsrc \left(\rsrc, t  - \frac{ \Ror}{c}\right)
\end{align}
where $\Ror$ is a shorthand notation for $\Ror = R(\rrec, \rsrc) = \|\rrec - \rsrc\|_{2}$ and $\inputsrc(\rsrc, t)$ is a source amplitude term (a mass flow rate per second, i.e. units of \si{\kilo \gram} s$^{-2}$) at a position $\rsrc$ within $\domain \setminus \bnd$. Without loss of generality, we assume the source to be a single point source, however, the treatment as follows can be straightforwardly generalized to a case of multiple point sources or a continuous source distribution. For this reason, we also drop the dependency of the source on $\rsrc$ in the following.


\section{An operator state-space representation}
\label{sec:OSS}
State-space models can be used to represent system of N first order differential equations. However in our case, we have not started from a differential equation, but rather an integral equation. Hence, the formulations we develop in this section are not conventional state-space models, and for that reason, we will refer to them as \textit{operator state-space models}. The main difference will be that as opposed to matrices and vectors, we will be using operators and functions in general. As a result, we need to consider the theory of functional analysis, but in fact we only need the basics, i.e., vectors become functions, matrices become operators, matrix multiplication becomes operator composition. Nevertheless, by still casting the problem in the general sense of state-space models, we pave the way to invoke fundamental theorems of state-space theory that can potentially lead to new perspectives on solving wave propagation problems with boundary conditions, such as in rooms, which is our primary interest. Additionally, this approach allows us to postpone discretization to a later stage and proceed in continuous time and continuous space, thus maintaining an exact connection to the physics. In this section, firstly, we will simplify the notation of the BIE as developed so far, then we will develop the operator state-space representation in the time-domain and the Laplace domain (i.e., both in continuous-time and continuous-space).

\subsection{A digestible notation}
\label{sec:OSS_td}

Even though \eqref{eq:total_soln} is valid for $\rrec \in \mathbb{R}^{3}$, we explicitly consider $p(\rrec, t)$ as a function with the domain of $p$, $\text{Dom}(p) = \domain \setminus \bnd$, hence $\weightbie(\rrec) = 4 \pi$. Substitution of \eqref{eq:KHIT_TD_GFsub_impedance}, \eqref{eq:pinc_td1}, and $\weightbie(\rrec) = 4 \pi$ in \eqref{eq:total_soln} results in
\begin{align}
	\label{eq:total_soln_br}
	p(\rrec, t) =&  \int_{\bnd}  \gbr   \frac{\partial \pbndry}{ \partial t} \left(\rbeta, t - \frac{\Rsr}{c}\right)      \hspace{0.1cm} d\bnd \nonumber \\
	&+  \int_{\bnd} \hbr \hspace{0.1cm} \pbndry \left(\rbeta, t - \frac{\Rsr}{c}\right)    d\bnd \hspace{0.1cm} \nonumber \\
	&+ \frac{1}{ \Ror} \hspace{0.1cm} \inputsrc\left(t  - \frac{ \Ror}{c}\right),
\end{align}
where
\begin{align}
	\label{eq:alpha_beta_br}
	\gbr &= \frac{1}{4\pi} \left( - \frac{\cos\thetabetar}{c\Rsr} +  \frac{\rho}{\Rsr Z(\rbeta)} \right) \\
	\hbr &=  - \frac{\cos\thetabetar}{4 \pi \Rsr^{2}} 
\end{align}

\noindent In order to evaluate \eqref{eq:total_soln_br}, both the pressure and the time-derivative of the pressure on the boundary are required. We can use the BIE once again to obtain an integral equation for the pressure on the boundary, i.e., \eqref{eq:total_soln} is also valid for $\rrec = $\rb, where $\rb$ is some position on the boundary. In this case, we consider the domain of the pressure on the boundary, $\text{Dom}(q) = \bnd$, so that $\weightbie(\rrec) = \weightbie(\rb) = 2 \pi$ (cfr. \eqref{eq:KHIT_scaling}). Hence analogous to  \eqref{eq:total_soln_br}, we can arrive to the following
\begin{align}
	\label{eq:total_soln_bb}
	\pbndry(\rb, t) =&  \int_{\bnd}  \gbb   \frac{\partial \pbndry}{ \partial t} \left(\rbeta, t - \frac{\Rbb}{c}\right)      \hspace{0.1cm} d\bnd \nonumber \\
	&+  \int_{\bnd} \hbb \hspace{0.1cm} \pbndry \left(\rbeta, t - \frac{\Rbb}{c}\right)    d\bnd \hspace{0.1cm} \nonumber \\
	&+ \frac{2}{\Ros} \hspace{0.1cm} \inputsrc \left(t  - \frac{ \Ros}{c}\right),
\end{align}
\begin{align}
	\label{eq:alpha_beta_bb}
	\gbb &= \frac{1}{2\pi} \left( - \frac{\cos\thetabb}{c\Rbb} +  \frac{\rho}{\Rbb Z(\rbeta)} \right) \\
	\label{eq:hbb1}
	\hbb &=  - \frac{\cos\thetabb}{2 \pi \Rbb^{2}} \\
	\Ros &= R(\rb, \rsrc) =  \|\rb - \rsrc\|_{2} \\
	\Rbb &= R(\rb, \rbeta) = \|\rb - \rbeta\|_{2} \\
	\cos \thetabb &= \theta (\rb, \rbeta) = \frac{(\rb -\rbeta)^{T}  \mathbf{n} } {\Rbb } 
\end{align}

\subsection{Time-Domain}

In the following, we propose an operator state-space representation where the pressure at the boundary, $\pbndry (\rb, t)$, is considered to be the internal state of the system. Using \eqref{eq:total_soln_bb} and \eqref{eq:total_soln_br}, we define the operator state-space representation for the BIE as

\begin{align}
	\label{eq:oss_td_state}
	\pbndry (\rb, t) &= \Asstd[\pbndry(\rb, t)](\rb, t) + \Bsstd[\inputsrc( t)](\rb,t),  \\
	\label{eq:oss_td_meas}
	p(\rrec, t) &= \Csstd[\pbndry(\rb, t)](\rrec, t) +   \Dsstd[\inputsrc( t)](\rrec,t)
\end{align}
where the integral operators are defined as
\begin{align}
	\label{eq:operators_td2}
	\Asstd [\pbndry(\rb, t)](\rb, t) =&  \int_{\bnd}  \gbb   \frac{\partial \pbndry}{ \partial t} \left(\rbeta, t - \frac{\Rbb}{c}\right)      \hspace{0.1cm} d\bnd  \nonumber \\
	&+  \int_{\bnd} \hbb \hspace{0.1cm} \pbndry  \left(\rbeta, t - \frac{\Rbb}{c}\right)    d\bnd , \hspace{0.1cm} \\
	\Bsstd [\inputsrc( t)](\rb,t) &= \frac{2}{\Ros} \hspace{0.1cm} \inputsrc \left( t  - \frac{ \Ros}{c}\right), \\
	\Csstd [\pbndry(\rb, t)](\rrec, t) =&  \int_{\bnd}  \gbr   \frac{\partial \pbndry}{ \partial t}  \left(\rbeta, t - \frac{\Rsr}{c}\right)      \hspace{0.1cm} d\bnd  \nonumber \\
	&+  \int_{\bnd} \hbr \hspace{0.1cm}  \pbndry  \left(\rbeta, t - \frac{\Rsr}{c}\right)    d\bnd ,  \\
	\Dsstd [\inputsrc( t)](\rrec, t) &= \frac{1}{ \Ror} \hspace{0.1cm} \inputsrc \left( t  - \frac{ \Ror}{c}\right).
\end{align}

\begin{figure}[t!]
	\centering
	
	\begin{subfigure}[b]{0.9\linewidth}
		\centering
		\includegraphics[width=\linewidth]{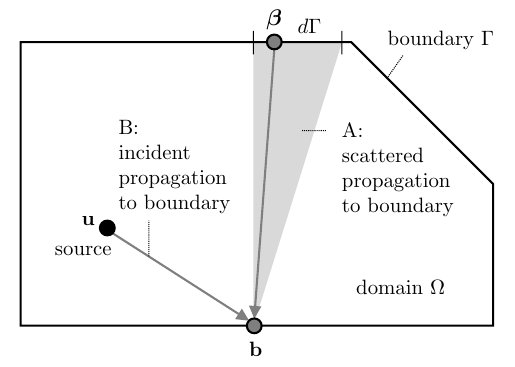}
		\caption{Wave propagation represented by the state equation.}
		\label{fig:room_state_eq}
	\end{subfigure}
	
	\begin{subfigure}[b]{0.9\linewidth}
		\centering
		\includegraphics[width=\linewidth]{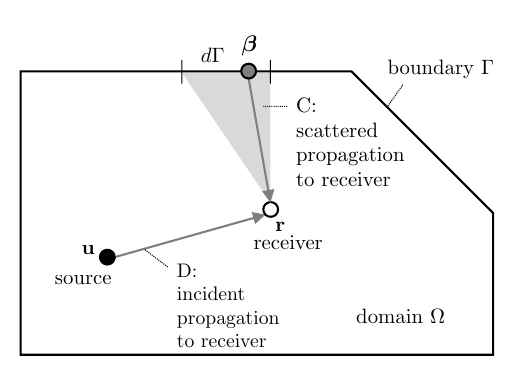}
		\caption{Wave propagation represented by the measurement equation.}
		\label{fig:room_meas_eq}
	\end{subfigure}
	
	\caption{State-space depiction of the boundary integral equation for the interior domain, $\domain$ enclosed by a boundary, $\bnd$. The state equation of (a) is used to compute the internal state, the pressure on the boundary, which is subsequently used in the measurement equation of (b) to obtain the pressure at the receiver. Generic symbols are used for the operators $A, B, C, D$ as this illustration holds for any of the operator state-space representations in  Sec. \ref{sec:OSS_td}, Sec. \ref{sec:OSS_lap}, or Sec. \ref{sec:spatialdiscretization}.}
	\label{fig:room_state_space}
\end{figure}

We interpret \eqref{eq:oss_td_state} as a so-called state equation, illustrated in Fig. \ref{fig:room_state_space} (a). The operator, $\Bsstd$, applies an amplitude scaling and a time delay to the input source, $\inputsrc(t)$, (corresponding to a propagation delay from the source to the boundary, i.e., $\Ros/c$) producing some initial pressure on the boundary, i.e., the result of this operation is a function of $(\rb, t)$. The state-transition operator, $\Asstd$, ``scatters'' the boundary pressure across the boundaries. Amplitude scalings ($\gbb$, $\hbb$) and time delays are applied to both the boundary pressure and its time derivative (corresponding to propagation delays from one boundary position to another, i.e., $\Rbb/c$), and integrated over $\rbeta$ on the boundary. The result of this operation is also a function of $(\rb, t)$ and accounts for how the boundary pressure, $\pbndry (\rb, t)$ is updated in this state equation. 

We interpret \eqref{eq:oss_td_meas} as a so-called measurement or observation equation, illustrated in Fig.~\ref{fig:room_state_space}~(b). The illustration in Fig.~\ref{fig:room_state_space}~(b) corresponds to that of Fig. \ref{fig:room_scene}, but explicitly shows how the propagation is represented by the respective operators from \eqref{eq:oss_td_meas}. The operator, $\Csstd$, applies amplitude scalings ($\gbr$, $\hbr$) and time delays to both the boundary pressure and its time derivative (corresponding to propagation delays from the boundary to the receiver, i.e., $\Rsr/c$) and the result is integrated over the boundary. This produces the scattered pressure contribution, $\pscat(\rrec, t)$ at the receiver. The incident pressure contribution, $\pinc(\rrec, t)$ at the receiver, is obtained via the operator $\Dsstd$, which applies an amplitude scaling and a time delay to the input source (corresponding to a propagation delay from the source to the receiver, i.e., $\Ror/c$).

\begin{figure}[t]
	\centering
	\includegraphics[width=0.9\linewidth]{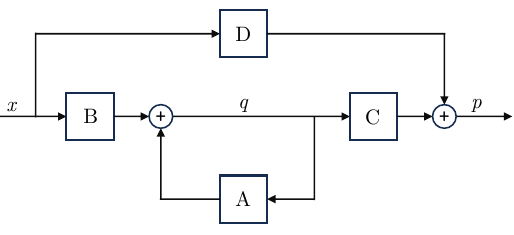}
	\caption{Block diagram representation of the state-space perspective of the boundary integral equation. $x$ represents the source term, $q$ the internal state, i.e., the pressure on the boundary, and $p$ the pressure at the receiver within $\domain \setminus \bnd$. Generic symbols are used for the operators $A, B, C, D$ as this illustration holds for any of the operator state-space representations in  Sec. \ref{sec:OSS_td}, Sec. \ref{sec:OSS_lap}, or Sec. \ref{sec:spatialdiscretization}.}
	\label{fig:statespaceblk}
\end{figure}

With the interpretation of the BIE from a state-space perspective via \eqref{eq:oss_td_state} and \eqref{eq:oss_td_meas}, we can represent the BIE using a block diagram representation as illustrated in Fig. \ref{fig:statespaceblk}. Generic symbols are used for the operators $A, B, C, D$ as this illustration holds not only for  \eqref{eq:oss_td_state} and \eqref{eq:oss_td_meas} (where $A$ corresponds to $\Asstd$, $B$ to $\Bsstd$, etc.), but also state-space representations we will develop in Sec. \ref{sec:OSS_lap} and Sec. \ref{sec:spatialdiscretization}. It is worthwhile to note that the scattering operation denoted by $A$ is responsible for the recursion within the system. Furthermore, changes to the source location only impacts the operation denoted by $B$ and $D$, and changes to the receiver location only impacts the operation denoted by $C$ and $D$.

\subsection{Laplace Domain}
\label{sec:OSS_lap}

Towards achieving a state-space representation without operators (see Sec. \ref{sec:spatialdiscretization}), we will firstly transform \eqref{eq:oss_td_state} and  \eqref{eq:oss_td_meas} into the Laplace domain. Let us recall the definition of the Laplace transform of a presumed causal function, $f(t)$ \cite{SignalsOppenheim}:
\begin{align}
	\label{eq:laplacetrans}
	L[f(t)](s) = \int_{0}^{\infty} f(t) e^{-st} dt,
\end{align}
where $L[.]$ is the Laplace operator, and $s = \sigma + j\omega$ is the Laplace variable. The Laplace transform properties for time-shifting and first-order time-derivative are given respectively as
\begin{align}
	\label{eq:laplace_shift}
	L[f(t-a)](s) &=  e^{-as} L[f(t)](s), \\
	\label{eq:laplace_deriv}
	L \left[ \frac{\partial f}{\partial t} (t)  \right](s) &=  s L[f(t)](s) - f(0), \hspace{0.5cm}
\end{align}
where $a > 0$, and $f(0)$ is an initial condition we assume to be zero. Applying these properties to \eqref{eq:total_soln_bb} and \eqref{eq:total_soln_br} results in the following 
\begin{align}
	\label{eq:integral_eqns_laplace}
	\pbndry (\rb, s) =&  \int_{\bnd}  (s \gbb + \hbb) e^{-s(\Rbb/c)}   \pbndry(\rbeta, s)       \hspace{0.1cm} d\bnd \nonumber \\
	& + \frac{2}{\Ros} \hspace{0.1cm} e^{-s(\Ros/c)} \inputsrc( s),
	&\\
	&\nonumber \\
	\label{eq:integral_eqns_laplace_obs}
	p(\rrec, s) =& \int_{\bnd}  (s \gbr + \hbr) e^{-s(\Rsr/c)}  \pbndry(\rb, s)      \hspace{0.1cm} d\bnd \nonumber \\
	&+ \frac{1}{ \Ror} e^{-s(\Ror/c)} \inputsrc( s),
\end{align}
from which we can make the following operator definitions
\begin{align}
	\label{eq:operators}
	A [\pbndry(\rb, s)](\rb, s) &= \int_{\bnd}  (s \gbb + \hbb) e^{-s(\Rbb/c)}   \pbndry(\rbeta, s)       \hspace{0.1cm} d\bnd \\
	\label{eq:Boperators}
	B [\inputsrc( s)](\rb,s) &= \frac{2}{\Ros} \hspace{0.1cm} e^{-s(\Ros/c)} \inputsrc( s) \\
	C [\pbndry(\rb, s)](\rrec, s) &=  \int_{\bnd}  (s \gbr + \hbr) e^{-s(\Rsr/c)}  \pbndry(\rb, s)      \hspace{0.1cm} d\bnd \\
	\label{eq:Doperators} 
	D [\inputsrc( s)](\rrec,s) &= \frac{1}{ \Ror} e^{-s(\Ror/c)}\inputsrc( s)
\end{align}

which allows us to express \eqref{eq:integral_eqns_laplace} and \eqref{eq:integral_eqns_laplace_obs} as an operator state-space model in the Laplace domain:
\begin{align}
	\label{eq:OSS-BIE-LD-stateeqn}
	\pbndry(\rb, s) &= A [\pbndry(\rb, s)](\rb, s) + B [\inputsrc( s)](\rb,s),  \\
	\label{eq:OSS-BIE-LD-measeqn}
	p(\rrec, s) &= C [\pbndry(\rb, s)](\rrec, s) +   D [\inputsrc( s)](\rrec,s).
\end{align}
Once again, the resemblance to conventional state-space models is apparent, but with operators and functions as opposed to matrices and vectors. We also take note again that the time-delays are incorporated into the definition of the operators.

\section{Spatial Discretization}
\label{sec:spatialdiscretization}

For practical applications, we aim to achieve a state-space representation that is free of both integral operators and time-shift operators.  As the Laplace transform converts the time-shift operator into a product, the problem reduces to discretizing integral operators in the Laplace domain, that is to spatial discretization of the boundary integral\footnote{For example, that the operators in \eqref{eq:Boperators} and \eqref{eq:Doperators} have already reduced to mere products. Note that discretization of the boundary integral could also be performed directly in time, but does not lead to an operator-free representation unless time is also discretized. For the purpose of this paper, we will not address time discretization and therefore work in the Laplace domain, which however could be mapped to the $z$-domain and hence discrete time if required, cf. also section V-A 1.}.

We approximate $\pbndry(\rb,s)$ as a sum of $N$ weighted shape or basis functions $\phi_\nu(\rb)$,
\begin{align}
	\pbndry(\rb,s) &= \sum_{\nu = 1}^N \pbndry_\nu(s) \phi_\nu(\rb) \label{eq:shape_functions}
\end{align}
where the support of $\phi_\nu(\rb)$ is denoted by $\Gamma_\nu \subseteq \Gamma$ and $\pbndry_\nu(s)$ is the corresponding weight.
We first consider discretization of the boundary pressure in the state equation.
Inserting into \eqref{eq:OSS-BIE-LD-stateeqn} yields
\begin{align}
	\pbndry(\rb, s) &= A \Bigl[\sum_{\nu} \pbndry_\nu(s) \phi_\nu(\rb)\Bigr](\rb, s) + B [\inputsrc( s)](\rb,s)\nonumber \\
	&= \sum_{\nu} A_\nu(\rb, s) \pbndry_\nu(s)  + B(\rb,s)\inputsrc(s)
	\label{eq:state_eq_shapefunctions}
\end{align}
where $A_\nu(\rb, s)$ and $B(\rb, s)$ are defined as
\begin{align}
	A_\nu(\rb, s) &= A [ \phi_\nu(\rb)](\rb, s) \nonumber \\
	&= \int_{\bnd_\nu}  \phi_\nu(\rbeta)  (s \gbb + \hbb) e^{-s(\Rbb/c)}   \hspace{0.1cm} d\bnd_\nu
	\label{eq:A_nu},\\
	B(\rb,s) &= \frac{2}{\Ros} \hspace{0.1cm} e^{-s(\Ros/c)},
\end{align}
and the summation commutes with integration and $\pbndry_\nu(s)$ can be taken out of the integral operator because it does not depend on the integration variable $\bnd_\nu$.
On the right-hand side of \eqref{eq:state_eq_shapefunctions} the boundary pressure is now expressed through the discrete weights $\pbndry_\nu(s)$, however, on the left-hand side it still remains a function of continuous space $\rb$.
This discrepancy can be resolved by a Galerkin method \cite{kirkup2019boundaryX}, which enforces that $\pbndry_{n}(s)$ satisfies 
\begin{align}
	\pbndry_{n}(s) =\int_{\Gamma_n} \tilde{\phi}_n(\rb)\pbndry(\rb, s)  \hspace{0.1cm} d\Gamma_n \label{eq:testing}
\end{align}
where we introduce the subscript $n$ instead of $\nu$ for later use and where  $\tilde{\phi}_n(\rb)$ is the so-called test function that is commonly chosen as  $\tilde{\phi}_n(\rb) = {\phi}_n(\rb)$. Note that \eqref{eq:testing} and \eqref{eq:shape_functions} can only hold exactly if $\tilde{\phi}_n(\rb) = {\phi}_n(\rb)$ form an orthonormal basis. Inserting \eqref{eq:state_eq_shapefunctions} in \eqref{eq:testing}, we obtain a set of $N$ equations
\begin{align}
	\pbndry_{n}(s) &=
	\sum_{\nu}A_{n,\nu}(s)\pbndry_\nu(s)
	+ B_{n}(s)\inputsrc(s),
	\label{eq:state_eq_testing}
\end{align}
with the scalars $A_{n,\nu}(s)$ and $B_{n}(s)$ defined as
\begin{align}
	A_{n,\nu}(s) =
	&\int_{\Gamma_n} \tilde{\phi}_n(\rb)    
	A_\nu(\rb, s) 
	\hspace{0.1cm} d\Gamma_n \nonumber\\
	= &\int_{\bnd_n} \int_{\bnd_\nu}\tilde{\phi}_n(\rb)\phi_\nu(\rbeta) \times \nonumber\\
	&(s \gbb + \hbb) e^{-s(\Rbb/c)}    \hspace{0.1cm} d\bnd_\nu
	\hspace{0.1cm} d\bnd_n, \label{eq:A_discrete}\\
	B_{n}(s) = 
	&\int_{\bnd_n} \tilde{\phi}_n(\rb) B(\rb,s) \hspace{0.1cm} d\bnd_n \nonumber\\
	= &\int_{\bnd_n} \tilde{\phi}_n(\rb) 
	\frac{2}{\Ros} \hspace{0.1cm} e^{-s(\Ros/c)} \hspace{0.1cm} d\bnd_n. \label{eq:B_discrete}
\end{align}
Note that the integral in \eqref{eq:A_discrete} exhibits singularities when $\rb = \rbeta$ due to the appearance of $\Rbb$ in $\gbb$ and $\hbb$ (see \eqref{eq:alpha_beta_bb} and \eqref{eq:hbb1}). The evaluation of such singular integrals has been well studied in the context of boundary element methods e.g., \cite[Ch. 8, sec. 4.3.2]{wu2000boundary}.

Let us now consider the discretization of the boundary pressure in the observation equation. 
Similarly to above, we insert \eqref{eq:shape_functions} into \eqref{eq:OSS-BIE-LD-measeqn}, which yields 
\begin{align}
	p(\rrec, s) &= C \Bigl[\sum_{\nu} \pbndry_\nu(s) \phi_\nu(\rb)\Bigr](\rrec, s) + D [\inputsrc( s)](\rrec,s),\nonumber \\
	&= \sum_{\nu} C_\nu(\rrec, s) \pbndry_\nu(s)  + D(\rrec, s)\inputsrc(s),
	\label{eq:obs_eq_shapefunctions}
\end{align}
with  $C_\nu(\rrec, s)$ and $D(\rrec, s)$ defined as
\begin{align}
	\label{eq:C_nu}
	C_\nu(\rrec, s) &= C [ \phi_\nu(\rb)](\rrec, s) \nonumber \\
	&= \int_{\bnd_\nu}  \phi_\nu(\rbeta)(s \gbr + \hbr) e^{-s(\Rsr/c)}    \hspace{0.1cm} d\bnd_\nu,\\
	D(\rrec, s) &= \frac{1}{ \Ror} e^{-s(\Ror/c)}.
\end{align}
Note that in case of the observation equation, a Galerkin method is not required since the left-hand side of \eqref{eq:obs_eq_shapefunctions} does not represent the pressure on the boundary but at the receiver, and therefore does not require discretization.

We now represent the set of equations \eqref{eq:state_eq_shapefunctions} and \eqref{eq:obs_eq_shapefunctions} by a matrix-vector expression considering a discrete set of receivers $\rrec_m$ with $m = 1,\, \dots,\, M$. 
We define

\begin{align}
\q (s) &= \begin{pmatrix} 
	q_1(s) & \cdots & q_N(s)
\end{pmatrix}^T,\\
\mathbf{p} (s) &= \begin{pmatrix} 
	p(\rrec_1, s) & \cdots & p(\rrec_M, s)
\end{pmatrix}^T,\\
\Assp (s) &= \begin{pmatrix} 
	A_{1,1}(s) & \cdots & A_{1,N}(s)\\
	\vdots & \ddots & \vdots\\
	A_{N,1}(s) & \cdots & A_{N,N}(s)\\
\end{pmatrix},\\
\Bssp (s) &= \begin{pmatrix} 
	B_{1}(s) \cdots B_{N}(s)
\end{pmatrix}^T,\\
	\Cssp (s) &= \begin{pmatrix} 
		C_1(\rrec_1, s)  & \cdots &C_N(\rrec_1, s)\\
		\vdots & \ddots & \vdots\\
		C_1(\rrec_M, s) & \cdots & C_N(\rrec_M, s)
	\end{pmatrix},\\
	\Dssp (s) &= \begin{pmatrix} 
		D(\rrec_1, s) \cdots D(\rrec_M, s)
	\end{pmatrix}^T.
\end{align}
Using these definitions, \eqref{eq:state_eq_shapefunctions} and \eqref{eq:obs_eq_shapefunctions} can be written as
\begin{align}
	\label{eq:OSS-BIE-LD-DiscSpace-stateeqn}
	\q (s) &= \Assp (s) \q (s) + \Bssp (s) x (s) \\
	\label{eq:OSS-BIE-LD-DiscSpace-measeqn}
	\mathbf{p} (s) &= \Cssp (s) \q (s)+ \Dssp (s) x (s).
\end{align}

\section{Discussion}
In this section, we will provide an extensive yet non-exhaustive overview of research problems in which the proposed framework can be adopted. We will point out relations of the \bioss{} framework to existing room acoustics models and illustrate how the \bioss{} framework may be used in the development of new estimation, inference, and control methods in room acoustics. We acknowledge that the various problems touched upon throughout this section require further research and, if successful, would comprise research contributions that go beyond the scope of this paper. The discussion will be organized by way of three types of problems involving room acoustics: forward problems (section \ref{subsec:existingmodels}) and inverse and control problems (section \ref{subsec:devnewmeth}).

\subsection{Positioning of existing room acoustics models in the proposed framework (forward problems)}\label{subsec:existingmodels}
The aim of this section is to show how existing room acoustics models fit into the proposed framework, with a particular focus on models developed to tackle forward problems, i.e., simulating or predicting the sound field at a set of receiver positions given the sound source signal and certain room properties (geometry, boundary conditions, etc.). This will facilitate the establishment of relations between existing models and may result in novel insights into existing models and their implementation or application. The discussion will be at a conceptual level, leaving a more rigorous treatment for future work. 

\begin{figure*}[t]
	\centering
	\includegraphics[width=0.7\linewidth]{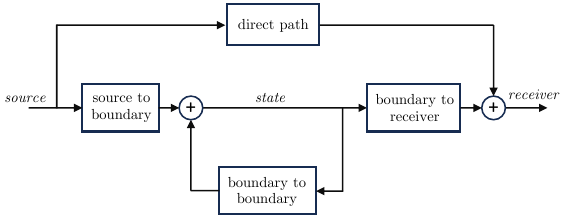}
	\caption{Block diagram representation of a generic room acoustics model featuring a feedback structure to represent the boundary-to-boundary scattering.}
	\label{fig:genericblk}
\end{figure*}

Conceptually, several room acoustics models can be represented by the block diagram depicted in Fig. \ref{fig:genericblk}. Next to a direct propagation of the source signal to the receiver, the source signal also generates a scattered sound field that is observed by the receiver. This scattered sound field is due to the propagation of the source signal to the boundary and the scattering of the boundary pressure to the room interior, including the receiver. At the heart of the block scheme, a positive feedback loop represents the recursive scattering of sound from the boundary to itself, which is typical for interior wave propagation problems. Comparing the block diagrams in Figs. \ref{fig:statespaceblk} and \ref{fig:genericblk}, it is obvious that this conceptual representation of room acoustics has a one-to-one correspondence to the proposed \bioss{} framework. 

While this correspondence may suggest that the \bioss{} framework only covers room acoustics models featuring a feedback structure, we will argue how the framework is in fact more widely applicable. The state equations in (\ref{eq:oss_td_state}), (\ref{eq:OSS-BIE-LD-stateeqn}) and (\ref{eq:OSS-BIE-LD-DiscSpace-stateeqn}) can be manipulated to yield equivalent non-recursive expressions
\begin{align}
	\label{eq:oss_td_state_manipulated}
	\pbndry (\rb, t) &= \left\{(\Isstd - \Asstd)^{-1} \circ \Bsstd \right\}[\inputsrc( t)](\rb,t),
\end{align}
in the time domain, with $\Isstd$ representing the identity operator and $\circ$ denoting operator composition, and
\begin{align}
	\label{eq:OSS-BIE-LD-stateeqn_manipulated}
	\pbndry(\rb, s) &= \left\{ (I - A)^{-1} \circ B \right\} [\inputsrc( s)](\rb,s),
\end{align}
in the Laplace domain, with $I$ representing the identity operator, or, after spatial discretization,
\begin{align}
	\label{eq:OSS-BIE-LD-DiscSpace-stateeqn_manipulated}
	\q (s) &= \bigl(\Issp - \Assp (s)\bigr)^{-1} \Bssp (s) x (s),
\end{align}
with $\Issp$ representing the identity matrix. Using the generic symbols introduced in the captions of Figs. \ref{fig:room_state_space} and \ref{fig:statespaceblk}, this can be written in a domain-agnostic manner as
\begin{align}
	\label{eq:OSS-BIE-generic-stateeqn_manipulated}
	\pbndry &= \bigl(\textrm{I} - \textrm{A}\bigr)^{-1} \textrm{B} \inputsrc,
\end{align}
where the notation has been simplified, in that $\textrm{A}\textrm{B}$ represents operator composition when $\textrm{A}$ and $\textrm{B}$ are operators or matrix multiplication when $\textrm{A}$ and $\textrm{B}$ are matrices. A sufficient but not necessary condition for the $(\textrm{I} - \textrm{A})$ operator to be invertible is that A is a continuous linear transformation over a Banach space and its norm is strictly less than one \cite[Th. 2.9]{sasane17}. Under these same assumptions,
the inverse of the $\left(\textrm{I} - \textrm{A}\right)$ operator corresponds to the Neumann series \cite[Th. 2.9]{sasane17}
\begin{align}
	\label{eq:Neumann_series_generic}
	\bigl(\textrm{I} - \textrm{A}\bigr)^{-1} &= \sum_{k=0}^{+\infty} \textrm{A}^k.
\end{align}
Here, $\textrm{A}^k$ is an iterated integral operator (in the continuous-space case) or a matrix power (in the discrete-space case), representing a $k$-fold repetition of the boundary-to-boundary scattering. The validity of the assumptions under which the $(\textrm{I} - \textrm{A})$ operator is invertible and the result in (\ref{eq:Neumann_series_generic}) holds, is dependent on the domain (time or Laplace) and the discretisation method used, and needs to be verified accordingly. 

Proceeding to substitute (\ref{eq:Neumann_series_generic}) into (\ref{eq:OSS-BIE-generic-stateeqn_manipulated}) then yields a non-recursive and inversion-free formulation of the state equation
\begin{align}
	\label{eq:OSS-BIE-generic-stateeqn_Neumann}
	\pbndry &= \sum_{k=0}^{+\infty} \textrm{A}^k \textrm{B} \inputsrc.
\end{align}

\begin{figure*}[t]
	\centering
	\includegraphics[width=0.95\linewidth]{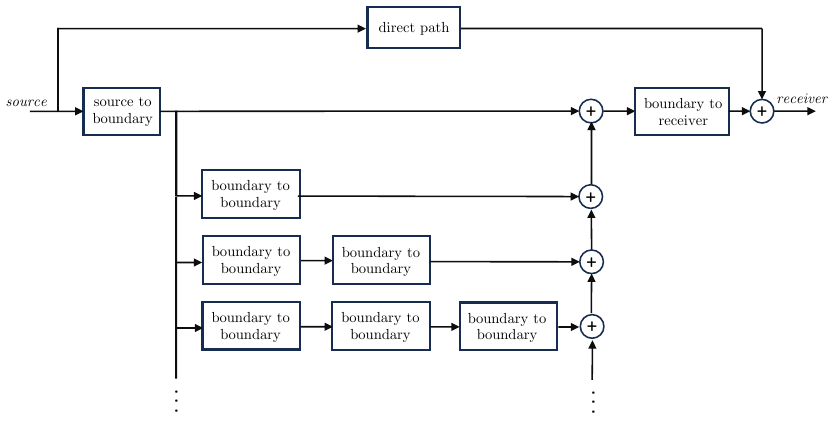}
	\caption{Block diagram representation of a generic room acoustics model featuring a parallel feedforward structure to represent the boundary-to-boundary scattering.}
	\label{fig:genericblk_parallel}
\end{figure*}

Combining this formulation with the \bioss{} measurement equation\footnote{Note that we can equivalently write the measurement equations (\ref{eq:oss_td_meas}), (\ref{eq:OSS-BIE-LD-measeqn}) and (\ref{eq:OSS-BIE-LD-DiscSpace-measeqn}) in a domain-agnostic manner.}, we obtain the following input-output relation for the \bioss{} framework
\begin{align}
	\label{eq:OSS-BIE-generic-measeqn}
	p &= \textrm{C} \pbndry + \textrm{D} \inputsrc \\
	\label{eq:OSS-BIE-generic-TF}
	&= \left( \textrm{C} \sum_{k=0}^{+\infty} \textrm{A}^k \textrm{B} + \textrm{D} \right) \inputsrc .
\end{align}
This input-output relation is depicted in Fig. \ref{fig:genericblk_parallel}, in which the feedback structure of Fig. \ref{fig:genericblk} has been expanded into a parallel feedforward structure of infinitely many increasing-order boundary-to-boundary scattering blocks, corresponding to the $\textrm{A}^k$ terms in (\ref{eq:OSS-BIE-generic-TF}).

As can be seen, the \bioss{} framework readily yields a transfer function interpretation of the BIE, where the transfer function $\textrm{T}$ relating the source signal $x$ to the pressure field $p = \textrm{T} x$ can be represented with or without inversion as
\begin{align}
	\textrm{T} = \textrm{C} \bigl(\textrm{I} - \textrm{A}\bigr)^{-1} \textrm{B} + \textrm{D} = \textrm{C} \sum_{k=0}^{+\infty} \textrm{A}^k \textrm{B} + \textrm{D} .
\end{align}

\subsubsection{Boundary Element Methods}

The frequency-domain boundary element method (FD-BEM) \cite{wu2000boundary} has a direct correspondence to the Laplace-domain state-space model in (\ref{eq:OSS-BIE-LD-DiscSpace-stateeqn})-(\ref{eq:OSS-BIE-LD-DiscSpace-measeqn}) by substituting $s = j\omega$. A large body of the FD-BEM literature has focused on the development of computationally efficient solvers for the linear system of equations resulting from (\ref{eq:OSS-BIE-LD-DiscSpace-stateeqn})-(\ref{eq:OSS-BIE-LD-DiscSpace-measeqn}), aiming to find the pressure field $p$ given the source distribution $x$. The application of the Neumann series result as in (\ref{eq:OSS-BIE-generic-stateeqn_Neumann}) allows for a computationally attractive approximation of the solution to (\ref{eq:OSS-BIE-LD-DiscSpace-stateeqn})-(\ref{eq:OSS-BIE-LD-DiscSpace-measeqn}), by truncating the series expansion to a finite order. This approximation strongly resembles a popular model order reduction technique for FD-BEM, known as Krylov subspace approximation \cite{sakuma08,panagiotopoulos20}.

The TD-BEM \cite{Hargreaves2007, wu2000boundary}, on the other hand, fits into the proposed framework in two ways. Firstly, starting from the time-domain state-space model in (\ref{eq:oss_td_state})-(\ref{eq:oss_td_meas}), a spatial basis expansion similar to (41) can be assumed for the space-time boundary pressure function $q(b,t)$, after which the time-domain coefficients $q_{\nu}(t)$ can be discretized. This is the common approach in TD-BEM, which is however known to result in unstable updating schemes \cite{Rynne1985, Smith01101990}. Secondly, starting from the Laplace-domain state-space model in (\ref{eq:OSS-BIE-LD-DiscSpace-stateeqn})-(\ref{eq:OSS-BIE-LD-DiscSpace-measeqn}), an appropriate $s$-to-$z$ mapping can be chosen, after which a time-discrete implementation can be obtained. This approach is reminiscent of the convolution quadrature method \cite{lubich1994multistep, banjai2014fast, sauter2017convolution}. Perhaps the most relevant open question concerning TD-BEM in relation to the proposed framework, is if the \bioss{} framework allows for a better analysis or solution to the TD-BEM stability problem. Firstly, stability results for state-space systems (e.g., Lyapunov stability \cite{bhatia70}) could potentially be extended to TD-BEM. Secondly, a feedback-free implementation resulting from the application of the Neumann series result as shown in Fig. \ref{fig:genericblk_parallel}, involving a truncation of the series expansion to a finite order, may pave the way towards a feedforward TD-BEM implementation, which is inherently stable. 
Note that the representation of the boundary-to-boundary scattering by means of a state-transition matrix has previously been suggested in TD-BEM literature \cite{Hargreaves2007}, yet adopting a definition of the state vector that differs from ours in that the state vector includes (time-discretised) retarded potentials.

\subsubsection{Delay networks}
Delay networks designed for artificial reverberation synthesis, more specifically the FDN \cite{jot1991digital} and SDN \cite{DeSena2015}, are based on a multi-channel feedback structure. In the SDN, the room boundary is sampled by defining scattering nodes at the walls, typically one node per wall \cite{DeSena2015}. The SDN feedback structure then consists of transfer function blocks linked by source-to-node connections, node-to-node connections, and node-to-microphone connections, in addition to a direct source-to-microphone connection. While this structure bears a strong resemblance to the block diagram in Fig. \ref{fig:genericblk}, the SDN state vector is defined differently from the \bioss{} state vector, in that pressure values originating from different nodes (referred to as wave variables) end up in different state variables. Nevertheless, it can be postulated that SDN corresponds to a particular instance of the \bioss{} framework. 
In other words, we anticipate that for every SDN satisfying certain design constraints, a corresponding $z$-domain model can be found that results from a Laplace-to-$z$-domain mapping of the discrete-space, Laplace-domain \bioss{} formulation in (\ref{eq:OSS-BIE-LD-DiscSpace-stateeqn})-(\ref{eq:OSS-BIE-LD-DiscSpace-measeqn}). A detailed treatment of the conditions under which this equivalence holds is outside the scope of this paper, but will be part of future work.

In contrast, for FDN, a formal correspondence to the \bioss{} framework cannot generally be established. Even though the FDN admits a state-space representation \cite{jot1991digital}, its parameters (\textit{i.e.}, delays and gain factors) are not constrained to correspond to a physically realizable interior wave propagation problem and therefore a corresponding model in the \bioss{} framework may not exist.

\subsubsection{Geometric models}
In geometric models such as the image source method (ISM) \cite{Allen1979} and the ray tracing (RT) method \cite{krokstad68}, boundary scattering is represented by specular reflections at the walls of the room. These models are non-recursive and inversion-free, whose structure matches with the general room acoustics model structure of Fig. \ref{fig:genericblk_parallel}, in that they start and end by modelling source-to-boundary propagation and boundary-to-boundary scattering, with a parallel structure of increasing-order boundary-to-boundary scattering operations. Whereas in the RT method, this increasing-order boundary-to-boundary scattering is modeled by tracing the reflection path of a sound ray along multiple walls, in the ISM it is modelled by virtually mirroring the room setup outside the physical room and considering the mirrored sources as image sources each producing a single reflection. The representation of the boundary-to-source scattering and boundary-to-boundary scattering operations in the ISM and RT methods rely on the physical laws of reflection. Whereas the geometric modelling of scattering by specular reflections is often considered an approximation, it has been shown in our previous work that for first-order reflections off a rigid and infinite planar wall, the boundary-to-source scattering operator (\textit{i.e.}, the C operator in the \bioss{} framework) is asymptotically equivalent to a specular reflection \cite{ali2023relating}. An extension of this work to show under which conditions an asymptotic equivalence can be established between the state-transition matrix (\textit{i.e.}, the A operator in the \bioss{} framework) and the (visible) higher-order specular reflections is currently being developed by the authors. 

\subsection{Development of new methods for estimation, inference and control in room acoustics}
\label{subsec:devnewmeth}
Whereas the previous section has focused on forward problems in room acoustics, i.e., problems involving the estimation of the sound field given a source term and boundary information, in this section we will illustrate how inverse and control problems can be tackled in the \bioss{} framework. 

Sound field reconstruction, sometimes also referred to as sound field or RIR interpolation, is an inverse problem that has received much attention in recent years in the context of room acoustics \cite{antonello17}. Sound field reconstruction methods are often based on a hybrid data-driven and physics-informed approach \cite{niebler22,karakonstantis24c,pezzoli23,tsunokini24,olivieri24,ma24,sundstrom24,karakonstantis23,fernandezgrande23,damiano25}. The \bioss{} framework provides an entirely new framework for such hybrid approaches. The \bioss{} operators $\{\textrm{A},\textrm{B},\textrm{C},\textrm{D} \}$ may be constructed from physical knowledge of the boundary geometry and impedance, and the source and receiver positions. Given a source term, these operators can be used to predict a state sequence and subsequently a pressure field as outlined above in the context of forward problems. However, in case only partial physical knowledge is available, the \bioss{} operators $\{\textrm{A},\textrm{B},\textrm{C},\textrm{D} \}$ may only be approximately constructed. This could for example be the case if boundary impedance values are unknown and the $\{\textrm{A},\textrm{C} \}$ operators are approximately constructed under the assumption of rigid boundary conditions or, similarly, if only a coarse approximation of the boundary geometry is known when constructing the $\{\textrm{A},\textrm{B},\textrm{C} \}$ operators. Despite the partial availability of physical knowledge in such cases, sound field reconstruction may still be achieved if a set of pressure field signals or RIRs can be measured at various receiver positions in the room. The approximate $\{\textrm{A},\textrm{B},\textrm{C},\textrm{D} \}$ operators can then be used jointly with the source term and receiver measurements to estimate a sequence of state vectors in a Kalman filtering framework, after which the state vector sequence can be used to estimate the sound field at receiver positions where no measurements are available, using $\{ \textrm{C},\textrm{D} \}$ operators appropriate for these receiver positions. The state-space structure of the \bioss{} framework suggests that it may be useful in this context to revisit the concept of observability for determining how many and at which receiver positions measurements need to be available to be able to estimate the state trajectory. A comprehensive treatment of this subject is left for future work, but it is reasonable to hypothesize that the operator observability matrix 
\begin{align}
	\mathcal{O} = \begin{bmatrix} \textrm{C} & \textrm{C} \textrm{A} & \textrm{C} \textrm{A}^2 & \ldots & \textrm{C} \textrm{A}^{N-1} \end{bmatrix}^T,
\end{align}
will play a crucial role in this context.

Even in case no physical knowledge of the room geometry, boundary impedance, and source position is available, the \bioss{} framework may still provide a suitable model structure for a fully data-driven approach to room acoustics inference problems. Assuming that a data set of source and receiver signals $\{ x, p \}$ is available, the \bioss{} operators $\{\textrm{A},\textrm{B},\textrm{C},\textrm{D} \}$ may be estimated by solving a state realization problem. In state realization problems for linear dynamical systems, the state-space model matrices are typically estimated from the so-called Markov parameters \cite{de2000minimal}, for which various algorithms exist \cite{ho66,moonen93,vanoverschee96}. Translating this concept to the \bioss{} framework, the Markov parameters would be defined as the set of operators 
\begin{align}
	\mathcal{M} = \{ \textrm{CB},\ \textrm{C} \textrm{A} \textrm{B},\ \textrm{C} \textrm{A}^2 \textrm{B},\ \ldots \}.
\end{align}
Notably, these parameters precisely correspond to the terms in the summation that describes the scattered field in the input-ouput relation (\ref{eq:OSS-BIE-generic-TF}). This observation suggests that, even if not obvious, the separation of the various orders of scattering components in a sound field or RIR measurement would potentially allow to identify the $\{\textrm{A},\textrm{B},\textrm{C},\textrm{D} \}$ operators from input-output data. As the $\textrm{A}$ operator, representing boundary-to-boundary scattering, is invariant to changes of the source or receiver position, a state-space model $\{\textrm{A},\textrm{B},\textrm{C},\textrm{D} \}$ identified for a particular set of source and receiver positions may be updated to different source and receiver positions within the same room by means of partial model updating \cite{de2000minimal}. In particular, in case of a source movement the operator pair $\{\textrm{B},\textrm{D} \}$ would need to be updated whereas in case of a receiver movement the operator pair $\{\textrm{C},\textrm{D} \}$ would need to be updated. Note that a state realization approach to model order reduction in room acoustics modelling was recently introduced in \cite{kujawski24}. In contrast to the \bioss{} framework proposed in this paper, the state-space model adopted in \cite{kujawski24} is not based on a physical model for space-time wave propagation. The Markov parameters are chosen to correspond to the discrete-time RIR coefficients, and no physical interpretation of the resulting state vector or state-space model matrices is given.

Due to the correspondence between the \bioss{} framework and existing room acoustics models, as elaborated in Sec. \ref{subsec:existingmodels}, the concept of fitting the \bioss{} model parameters to sound field or RIR measurements can also be extended to other models. This concept is not entirely new as it has been applied to fit the model parameters of the FDN to RIR measurements \cite{shen20,lyster22,bona22,ibnyahya22}. The resulting FDN is however not suitable for accurate sound field prediction, as the fitting process targets perceptual matching rather than physical matching. Attempts to fit the ISM model parameters to RIR measurements have not been successful to date \cite{sprunck2025}.

Finally, also the problem of sound field control can be cast into the \bioss{} framework. In this context, a key question is how many (and at which positions) loudspeakers (secondary sources) and microphones (control sensors) need to be deployed to achieve a certain target performance \cite{koyama20}. In state-space control, this question is answered by using the notion of controllability. Similarly to what has been stated above regarding observability, we may hypothesize that the operator controllability matrix 
\begin{align}
	\mathcal{C} = \begin{bmatrix} \textrm{B} & \textrm{A} \textrm{B}  & \textrm{A}^2 \textrm{B} & \ldots & \textrm{A}^{N-1} \textrm{B} \end{bmatrix}^T,
\end{align}
is relevant to consider in this context, however leaving a detailed treatment for future work.

\section{Conclusion}

In this paper, the boundary integral operator state-space (\bioss{}) framework for room acoustics modelling has been proposed, which provides a state-space representation of the boundary integral equation. Various instances of the framework have been developed, in the time and frequency domain, as well as in continuous and discrete space. While being rooted in linear acoustics theory, the \bioss{} framework provides a fresh and potentially appealing signals and systems perspective to room acoustics modelling, which opens various directions for future research. Equivalences between the feedback and parallel feedforward structures within the \bioss{} framework and existing room acoustics models have been pointed out, which pave the way for establishing relations between existing models and possibly developing new room acoustics models. State-space modelling concepts such as observability, controllability, and state realization can potentially be applied to the \bioss{} framework, creating new perspectives for solving room acoustics inference and control problems.

\funding

This research work was carried out in the frame of KU Leuven Internal Funds C14/21/075 and C3/23/056, FWO project G0A0424N and EPSRC project EP/V002554/1. The research leading to these results has received funding from the European Research Council under the European Union's Horizon 2020 research and innovation program / ERC Consolidator Grant: SONORA (no. 773268). This paper reflects only the authors' views and the Union is not liable for any use that may be made of the contained information.


\bibliographystyle{unsrt}
\bibliography{sonora_biblio_trimmed}

@String{AA = {Acta Acust.}}

@String{APMag = {IEEE Antennas Propagat. Mag.}}

@String{CACM = {Commun. ACM}}

@String{EM = {Electromagnetics}}

@String{FrontSP = {Front. Signal Process.}}

@String{JAES = {J. Audio Eng. Soc.}}

@String{JAM = {IMA J. Appl. Math.}}

@String{JASA = {J. Acoust. Soc. Amer.}}

@String{JASMP = {EURASIP J. Audio, Speech, Music Process.}}

@String{JCAM = {J. Comput. Appl. Math.}}

@String{JCP = {J. Comput. Phys.}}

@String{JSV = {J. Sound Vib.}}

@String{NM = {Numerische Mathematik}}

@String{PNAS = {Proc. Natl. Acad. Sci.}}

@String{RegTech = {Regelungstechnik}}

@String{SP = {Signal Processing}}

@String{TransAC = {IEEE Trans. Autom. Control}}

@String{TransAP = {IEEE Trans. Antennas Propag.}}

@String{TransASLP = {IEEE Trans. Audio Speech Lang. Process.}}

@String{TransASLP_ACM = {IEEE/ACM Trans. Audio Speech Lang. Process.}}

@String{TransSAP = {IEEE Trans. Speech Audio Process.}}

@String{TransSP = {IEEE Trans. Signal Process.}}

@String{AES90Conv = {Preprints AES 90th Convention}}

@String{AES122Conv = {Preprints AES 122nd Convention}}

@String{AES153Conv = {Preprints AES 153rd Convention}}

@String{AM20 = {Proc. 15th Int. Audio Mostly Conf. (AM '20)}}

@String{AM22 = {Proc. 17th Int. Audio Mostly Conf. (AM '22)}}

@String{DAFX24 = {Proc. 27th Int. Conf. Digital Audio Effects (DAFx '24)}}

@String{EAASA09 = {Proc. 2009 EAA Symp. Auralization}}

@String{EUSIPCO10 = {Proc. 18th European Signal Process. Conf. (EUSIPCO '10)}}

@String{EUSIPCO25 = {Proc. 31st European Signal Process. Conf. (EUSIPCO '25)}}

@String{FA20 = {Proc. 9th Conv. Eur. Acoust. Assoc. (Forum Acusticum '20)}}

@String{FA23 = {Proc. 10th Conv. Eur. Acoust. Assoc. (Forum Acusticum '23)}}

@String{ICA22 = {Proc. 24th Int. Congr. Acoust. (ICA '22)}}

@String{INTERNOISE22 = {Proc. 51st Int. Congress \& Exposition Noise Control Eng. (INTER-NOISE '22)}}

@String{IWAENC24 = {Proc. 2024 Int. Workshop Acoustic Signal Enhancement (IWAENC '24)}}

@String{SMC22 = {Proc. 19th Sound Music Comput. Conf. (SMC '22)}}

@String{VR19 = {Proc. 2019 IEEE Conf. Virtual Reality 3D User Interfaces (VR '19)}}

@article{de2000minimal,
  title={Minimal state-space realization in linear system theory: an overview},
  author={B. De Schutter},
  journal=JCAM,
  volume={121},
  number={1--2},
  pages={331--354},
  year={2000},
  publisher={Elsevier}
}

@article{banjai2014fast,
  title={Fast convolution quadrature for the wave equation in three dimensions},
  author={L. Banjai and M. Kachanovska},
  journal=JCP,
  volume={279},
  pages={103--126},
  year={2014},
  publisher={Elsevier}
}

@article{sauter2017convolution,
  title={Convolution quadrature for the wave equation with impedance boundary conditions},
  author={S. A. Sauter and M. Schanz},
  journal=JCP,
  volume={334},
  pages={442--459},
  year={2017},
  publisher={Elsevier}
}

@article{lubich1994multistep,
  title={On the multistep time discretization of linear initial-boundary value problems and their boundary integral equations},
  author={C. Lubich},
  journal=NM,
  volume={67},
  number={3},
  pages={365--389},
  year={1994},
  publisher={Springer}
}

@article{ergin2002plane,
  title={The plane-wave time-domain algorithm for the fast analysis of transient wave phenomena},
  author={A. A. Ergin and B. Shanker and E. Michielssen},
  journal=APMag,
  volume={41},
  number={4},
  pages={39--52},
  year={2002},
  publisher={IEEE}
}

@article{santo2024feedback,
  title={Optimizing tiny colorless Feedback Delay Networks},
  author={G. Dal Santo and K. Prawda and S. J. Schlecht and V. V{\"a}lim{\"a}ki},
  journal=JASMP,
  year =	{2024},
  volume =	{2024},
  month =	{Article No. 13, 15 pages,}
}

@article{siltanen2007room,
  title={The room acoustic rendering equation},
  author={S. Siltanen and T. Lokki and S. Kiminki and L. Savioja},
  journal=JASA,
  volume={122},
  number={3},
  pages={1624--1635},
  year={2007},
  publisher={AIP Publishing}
}

@article{scerbo2024modartmodaldecompositionacoustic,
      title={Modeling nonuniform energy decay through the Modal Decomposition of Acoustic Radiance Transfer ({MoD-ART})}, 
      author={M. Scerbo and S. J. Schlecht and R. Ali and L. Savioja and E. De Sena},
      journal=TransASLP,
      OPTvolume={122},
      OPTnumber={3},
      OPTpages={1624--1635},
      year={2025}
}

@article{kirkup2019boundaryX,
  title={The boundary element method in acoustics: A survey},
  author={Kirkup, Stephen},
  journal={Applied Sciences},
  volume={9},
  number={8},
  pages={1642},
  year={2019},
  publisher={MDPI}
}

@inproceedings{ali2023relating,
  title={Relating wave-based and geometric acoustics using a stationary phase approximation},
  author={R. Ali and T. Dietzen and M. Scerbo and L. Vinceslas and T. van Waterschoot and De Sena, E.},
  booktitle=FA23,
  year={2023},
  OPTorganization={European Acoustics Association}
}

@article{raghuvanshi2007real,
  title={Real-time sound synthesis and propagation for games},
  author={N. Raghuvanshi and C. Lauterbach and A. Chandak and D. Manocha and M. C. Lin},
  journal=CACM,
  volume={50},
  number={7},
  pages={66--73},
  year={2007},
  publisher={ACM New York, NY, USA}
}

@inproceedings{kim2019immersive,
  title={Immersive spatial audio reproduction for {VR/AR} using room acoustic modelling from 360 images},
  author={H. Kim and L. Remaggi and P. J. B. Jackson and A. Hilton},
  booktitle=VR19,
  pages={120--126},
  year={2019},
  OPTorganization={IEEE}
}

@article{Vorlander_2013, title={Computer simulations in room acoustics: Concepts and uncertainties}, volume={133}, ISSN={0001-4966}, DOI={10.1121/1.4788978}, number={3}, journal=JASA, author={Vorländer, M.}, year={2013}, month=mar, pages={1203–1213} }

@book{vorlander2020auralization,
  title={Auralization},
  author={M. Vorl{\"a}nder},
  year={2020},
  publisher={Springer}
}

@book{long2005architectural,
  title={Architectural acoustics},
  author={M. Long},
  year={2005},
  publisher={Elsevier}
}

@article{Rynne1985,
    author = {B. P. Rynne},
    title = {Stability and Convergence of Time Marching Methods in Scattering Problems},
    journal = JAM,
    volume = {35},
    number = {3},
    pages = {297--310},
    year = {1985},
    month = Nov
}

@article{Dokmanic2013,
author = {I. Dokmanić  and R. Parhizkar and A. Walther  and Y. M. Lu  and M. Vetterli},
title = {Acoustic echoes reveal room shape},
journal = PNAS,
volume = {110},
number = {30},
pages = {12186--12191},
year = {2013}
}

@ARTICLE{Antonello19,
  author={N. Antonello and De Sena, E. and M. Moonen and P. A. Naylor and T. van Waterschoot},
  journal=TransASLP_ACM, 
  title={Joint Acoustic Localization and Dereverberation Through Plane Wave Decomposition and Sparse Regularization}, 
  year={2019},
  volume={27},
  number={12},
  pages={1893--1905}
}

@ARTICLE{Sadigh1992,
  author={A. Sadigh and E. Arvas},
  journal=TransAP, 
  title={Treating the instabilities in marching-on-in-time method from a different perspective (electromagnetic scattering)}, 
  year={1993},
  volume={41},
  number={12},
  pages={1695--1702},
  doi={10.1109/8.273314}}

@article{Smith01101990,
author = {P. D. Smith},
title = {Instabilities in Time Marching Methods for Scattering: Cause and Rectification},
journal = EM,
volume = {10},
number = {4},
pages = {439--451},
year = {1990},
publisher = {Taylor \& Francis},
}

@inproceedings{enzner2010bayesian,
  title={Bayesian inference model for applications of time-varying acoustic system identification},
  author={G. Enzner},
  booktitle=EUSIPCO10,
  pages={2126--2130},
  year={2010}
}

@article{macwilliam2024state,
  title={State-space estimation of spatially dynamic room impulse responses using a room acoustic model-based prior},
  author={K. MacWilliam and T. Dietzen and R. Ali and T. van Waterschoot},
  journal=FrontSP,
  volume={4},
  OPTpages={1426082},
  year={2024},
  publisher={Frontiers Media SA}
}

@article{Mezza2024,
	author = {A. I. Mezza and R. Giampiccolo and De Sena, E. and A. Bernardini},
	journal = JASMP,
	title = {Data-driven room acoustic modeling via differentiable feedback delay networks with learnable delay lines},
	year = {2024},
  	volume =	{2024},
  	month =	{Article No. 51, 20 pages,}
}

@book{jarrett2017theory,
  title={Theory and applications of spherical microphone array processing},
  author={D. P. Jarrett and E. A. P. Habets and P. A. Naylor},
  OPTvolume={9},
  year={2017},
  publisher={Springer}
}

@article{pinto2010space,
  title={Space-time-frequency processing of acoustic wave fields: Theory, algorithms, and applications},
  author={F. Pinto and M. Vetterli},
  journal=TransSP,
  volume={58},
  number={9},
  pages={4608--4620},
  year={2010},
  publisher={IEEE}
}

@article{berkhout1997array,
  title={Array technology for acoustic wave field analysis in enclosures},
  author={A. J. Berkhout and D. de Vries and J. J. Sonke},
  journal=JASA,
  volume={102},
  number={5},
  pages={2757--2770},
  year={1997},
  publisher={Acoustical Society of America}
}

@article{tervo2013spatial,
  title={Spatial decomposition method for room impulse responses},
  author={S. Tervo and J. P{\"a}tynen and A. Kuusinen and T. Lokki},
  journal=JAES,
  volume={61},
  number={1/2},
  pages={17--28},
  year={2013},
  publisher={Audio Engineering Society}
}

@article{van2008optimally,
  title={Optimally regularized adaptive filtering algorithms for room acoustic signal enhancement},
  author={T. van Waterschoot and G. Rombouts and M. Moonen},
  journal=SP,
  volume={88},
  number={3},
  pages={594--611},
  year={2008},
  publisher={Elsevier}
}

@article{funkhouser2004beam,
  title={A beam tracing method for interactive architectural acoustics},
  author={T. Funkhouser and N. Tsingos and I. Carlbom and G. Elko and M. Sondhi and J. E. West and G. Pingali and P. Min and A. Ngan},
  journal=JASA,
  volume={115},
  number={2},
  pages={739--756},
  year={2004},
  publisher={Acoustical Society of America}
}

@article{krokstad1968calculating,
  title={Calculating the acoustical room response by the use of a ray tracing technique},
  author={A. Krokstad and S. Strom and S. S{\o}rsdal},
  journal=JSV,
  volume={8},
  number={1},
  pages={118--125},
  year={1968},
  publisher={Elsevier}
}

@article{ramo2014high,
  title={High-precision parallel graphic equalizer},
  author={J. R{\"a}m{\"o} and V. V{\"a}lim{\"a}ki and B. Bank},
  journal=TransASLP_ACM,
  volume={22},
  number={12},
  pages={1894--1904},
  year={2014},
}

@article{vairetti2017scalable,
  title={A scalable algorithm for physically motivated and sparse approximation of room impulse responses with orthonormal basis functions},
  author={G. Vairetti and De Sena, E. and M. Catrysse and S. H. Jensen and M. Moonen and T. van Waterschoot},
  journal=TransASLP_ACM,
  volume={25},
  number={7},
  pages={1547--1561},
  year={2017},
}

@article{haneda1994common,
  title={Common acoustical pole and zero modeling of room transfer functions},
  author={Y. Haneda and S. Makino and Y. Kaneda},
  journal=TransSAP,
  volume={2},
  number={2},
  pages={320--328},
  year={1994},
}

@article{merimaa2005spatial,
  title={Spatial impulse response rendering {I}: Analysis and synthesis},
  author={J. Merimaa and V. Pulkki},
  journal=JAES,
  volume={53},
  number={12},
  pages={1115--1127},
  year={2005},
}

@inproceedings{scerbo2024common,
  title={A Common-Slopes Late Reverberation Model Based on Acoustic Radiance Transfer},
  author={M. Scerbo and S. Schlecht and R. Ali and L. Savioja and De Sena, E.},
  booktitle=DAFX24,
  pages={270--277},
  year={2024},
  OPTorganization={University of Surrey}
}

@book{willems1997introduction,
  title={Introduction to mathematical systems theory: a behavioral approach},
  author={J. W. Polderman and J. C. Willems},
  OPTvolume={26},
  year={1997},
  publisher={Springer Science \& Business Media}
}

@inproceedings{jeong2022design,
  title={Design, simulation, and virtual prototyping of room acoustics: Challenges and opportunities},
  author={C.-H. Jeong},
  booktitle=ICA22,
  year={2022}
}

@book{Oppenheim2009,
	author = {A. V. Oppenheim and R. W. Schafer},
	title = {Discrete-Time Signal Processing},
	year = {2009},
	publisher = {Prentice Hall},
	OPTaddress = {USA},
	edition = {3rd}
}

@inproceedings{farina2007advancements,
  title={Advancements in impulse response measurements by sine sweeps},
  author={A. Farina},
  booktitle=AES122Conv,
  year={2007},
  note = {{AES} Preprint 7121},
  OPTorganization={Audio Engineering Society}
}

@article{Savioja2015,
    author = {L. Savioja and U. P. Svensson},
    title = {Overview of geometrical room acoustic modeling techniques},
    journal = JASA,
    volume = {138},
    number = {2},
    pages = {708--730},
    year = {2015},
    month = Aug
}

@book{bergman2018computational,
  title={Computational acoustics: theory and implementation},
  author={D. R. Bergman},
  year={2018},
  publisher={John Wiley \& Sons}
}

@book{SignalsOppenheim,
author = {A. V. Oppenheim and A. S. Willsky and S. H. Nawab},
title = {Signals \& systems},
year = {1996},
publisher = {Prentice-Hall},
address = {USA},
edition = {2nd}
}

@inproceedings{aretz2009combined,
  title={Combined broadband impulse responses using {FEM} and hybrid ray-based methods},
  author={M. Aretz and R. N{\"o}then and M. Vorl{\"a}nder and D. Schr{\"o}der},
  booktitle=EAASA09,
  pages={1--6},
  year={2009}
}

@article{Wittebol2024,
title = {A hybrid room acoustic modeling approach combining image source, acoustic diffusion equation, and time-domain discontinuous Galerkin methods},
journal = AA,
volume = {223},
OPTpages = {110068},
year = {2024},
issn = {0003-682X},
author = {W. Wittebol and H. Wang and M. Hornikx and P. Calamia}
}

@phdthesis{hamilton2016finite,
    title={Finite difference and finite volume methods for wave-based modelling of room acoustics},
    author={B. Hamilton},
    year={2016},
    school={University of Edinburgh},
    address = {Edinburgh, Scotland, UK}
}

@ARTICLE{Valimaki50yrs,
	author={V. V\"alim\"aki and J. D. Parker and L. Savioja and J. O. Smith and J. S. Abel},
	journal=TransASLP,
	title={Fifty Years of Artificial Reverberation}, 
	year={2012},
	volume={20},
	number={5},
	pages={1421--1448},
	doi={10.1109/TASL.2012.2189567}
}

@article{jot1991digital,
	author={J.-M. Jot and A. Chaigne},
	journal=AES90Conv,
	title={Digital Delay Networks for Designing Artificial Reverberators},
	year={1991},
	month=Feb,
	note = {{AES} Preprint 3030}
}

@book{wu2000boundary,
	title={Boundary Element Acoustics: Fundamentals and Computer Codes},
	author={T. W. Wu},
	isbn={9781853125706},
	lccn={99069081},
	series={Advances in boundary elements series},
	url={https://books.google.be/books?id=W75rtAEACAAJ},
	year={2000},
	publisher={WIT}
}

@book{Hoskins2011,
	title = {Delta Functions},
	author = {R. F. Hoskins},
	publisher = {Woodhead Publishing},
	edition = {2nd},
	year = {2011}
}

@book{kuttruff2009room,
	title={Room Acoustics},
	author={H. Kuttruff},
	isbn={9780203876374},
	lccn={2008051863},
	year={2009},
	edition = {5th},
	publisher={Taylor \& Francis}
}

@Book{Pierce81,
	author = {A. D. Pierce},
	title = {Acoustics: An introduction to its physical principles and applications},
	isbn = { 0070499616 0070499624 },
	publisher = {McGraw-Hill},
	pages = { xxiii, 642 p. : },
	year = { 1981 },
	type = { Book },
	language = { English },
	subjects = { Sound. },
}

@book{jacobsen2013,
	title={Fundamentals of General Linear Acoustics},
	author={F. Jacobsen and P. M. Juhl},
	isbn={9781118636176},
	lccn={2013016030},
	year={2013},
	publisher={Wiley}
}

@article{Allen1979,
	author = {J. B. Allen and D. A. Berkley},
	doi = {10.1121/1.382599},
	issn = {NA},
	journal = JASA,
	number = {4},
	pages = {943--950},
	title = {Image method for efficiently simulating small-room acoustics},
	volume = {65},
	year = {1979}
}

@ARTICLE{desena2015,
	AUTHOR = {De Sena, E. and H. Hac{\i}habibo\u{g}lu and Z. Cvetković and J. O. Smith},
	TITLE = {Efficient synthesis of room acoustics via scattering delay networks},
	JOURNAL = TransASLP_ACM,
	VOLUME = {23},
	NUMBER = {9},
	PAGES = {1478-1492},
	YEAR = {2015}
}

@PhdThesis{Hargreaves2007,
	OPTmonth = Apr,
	title = {Time domain boundary element method for room acoustics},
	school = {University of Salford},
	author = {J. A. Hargreaves},
	year = {2007},
	address = {Salford, UK},
	url = {http://usir.salford.ac.uk/id/eprint/16604/}
}

@book{cox2016acoustic,
	title={Acoustic absorbers and diffusers: theory, design and application},
	author={T. Cox and P. d’Antonio},
	year={2016},
	publisher={CRC Press}
}

@INCOLLECTION{sakuma08,
	AUTHOR = {T. Sakuma and S. Schneider and Y. Yasuda},
	TITLE = {Fast solution methods},
	BOOKTITLE = {Computational Acoustics of Noise Propagation in Fluids - Finite and Boundary Element Methods},
	EDITOR = {S. Marburg and B. Nolte},
	PUBLISHER = {Springer},
	YEAR = {2008}
}

@inproceedings{panagiotopoulos20,
  title={A two-step reduction method for acoustic {BEM} systems},
  author={D. Panagiotopoulos and E. Deckers and W. Desmet},
  booktitle=FA20,
  year={2020},
  pages = {2291--2298},
  OPTorganization={European Acoustics Association}
}

@book{sasane17,
	title={A friendly approach to functional analysis},
	author={A. Sasane},
	year={2017},
	publisher={World Scientific}
}

@book{bhatia70,
	title={Stability theory of dynamical systems},
	author={N. P. Bhatia and G. P. Szeg\"o},
	year={1970},
	publisher={Springer}
}

@ARTICLE{krokstad68,
	AUTHOR = {A. Krokstad and S. Strøm and S. Sørsdal},
	TITLE = {Calculating the acoustical room response by the use of a ray tracing technique},
	JOURNAL = JSV,
	VOLUME = {8},
	NUMBER = {1},
	PAGES = {118-125},
	YEAR = {1968}
}

@ARTICLE{antonello17,
	AUTHOR = {N. Antonello and others},
	TITLE = {Room impulse response interpolation using a sparse spatio-temporal representation of the sound field},
	JOURNAL = TransASLP_ACM,
	VOLUME = {25},
	NUMBER = {10},
	PAGES = {1929-1941},
	YEAR = {2017}
}

@INPROCEEDINGS{niebler22,
	AUTHOR = {K. Niebler and P. Bonnaire and N. A. K. Doan and C. F. Silva},
	TITLE = {Towards reconstruction of acoustic fields via physics-informed neural networks},
	BOOKTITLE = INTERNOISE22,
	ADDRESS = {Glasgow, Scotland, UK},
	VOLUME = {265},
	YEAR = {2022}
}

@ARTICLE{karakonstantis24c,
	AUTHOR = {X. Karakonstantis and D. Caviedes-Nozal and A. Richard and E. Fernandez-Grande},
	TITLE = {Room impulse response reconstruction with physics-informed deep learning},
	JOURNAL = JASA,
	VOLUME = {155},
	NUMBER = {2},
	PAGES = {1048-1059},
	YEAR = {2024}
}

@INPROCEEDINGS{pezzoli23,
	AUTHOR = {M. Pezzoli and F. Antonacci and A. Sarti},
	TITLE = {Implicit neural representation with physics-informed neural networks for the reconstruction of the early part of room impulse responses},
	BOOKTITLE = FA23,
	ADDRESS = {Turin, Italy},
	YEAR = {2023}
}

@ARTICLE{tsunokini24,
	AUTHOR = {I. Tsunokini and G. Sato and Y. Ikeda and Y. Oikawa},
	TITLE = {Spatial extrapolation of early room impulse responses with noise-robust physics-informed neural network},
	JOURNAL = {IEICE Trans. Fundam. Electron. Commun. Comput. Sci., Article No. 2024EAL2015},
	YEAR = {2024}
}

@ARTICLE{olivieri24,
	AUTHOR = {M. Olivieri and X. Karakonstantis and M. Pezzoli and F. Antonacci and A. Sarti and E. Fernandez-Grande},
	TITLE = {Physics-informed neural network for volumetric sound field reconstruction of speech signals},
	JOURNAL = JASMP,
	VOLUME = {2024},
	NOTE = {Article No. 42},
	YEAR = {2024}
}

@ARTICLE{ma24,
	AUTHOR = {F. Ma and S. Zhao and I. S. Burnett},
	TITLE = {Sound field reconstruction using a compact acoustics-informed neural network},
	JOURNAL = JASA,
	VOLUME = {156},
	NUMBER = {3},
	PAGES = {2009-2021},
	YEAR = {2024}
}

@INPROCEEDINGS{sundstrom24,
	AUTHOR = {D. Sundström and S. Koyama and A. Jakobsson},
	TITLE = {Sound field estimation using deep kernel learning regularized by the wave equation},
	BOOKTITLE = IWAENC24,
	ADDRESS = {Aalborg, Denmark},
	PAGES = {319-323},
	YEAR = {2024}
}

@ARTICLE{karakonstantis23,
	AUTHOR = {X. Karakonstantis and E. Fernandez-Grande},
	TITLE = {Generative adversarial networks with physical sound field priors},
	JOURNAL = JASA,
	VOLUME = {154},
	NUMBER = {2},
	PAGES = {1226-1238},
	YEAR = {2023}
}

@ARTICLE{fernandezgrande23,
	AUTHOR = {E. Fernandez-Grande and X. Karakonstantis and D. Caviedes-Nozal and P. Gerstoft},
	TITLE = {Generative models for sound field reconstruction},
	JOURNAL = JASA,
	VOLUME = {153},
	NUMBER = {2},
	PAGES = {1179-1190},
	YEAR = {2023}
}

@INPROCEEDINGS{damiano25,
	AUTHOR = {S. Damiano and T. van Waterschoot},
	TITLE = {Sound field reconstruction using physics-informed boundary integral networks},
	BOOKTITLE = EUSIPCO25,
	ADDRESS = {Palermo, Italy},
	PAGES = {76-80},
	YEAR = {2025}
}

@ARTICLE{ho66,
	AUTHOR = {B. L. Ho and R. E. Kalman},
	TITLE = {Effective construction of linear, state-variable models from input-output functions},
	JOURNAL = RegTech,
	VOLUME = {14},
	NUMBER = {12},
	PAGES = {545-548},
	YEAR = {1966}
}

@ARTICLE{moonen93,
	AUTHOR = {M. Moonen and J. Ramos},
	TITLE = {A subspace algorithm for balanced state space system identification},
	JOURNAL = TransAC,
	VOLUME = {38},
	NUMBER = {11},
	PAGES = {1727-1729},
	YEAR = {1993}
}

@book{vanoverschee96,
  title={Subspace identification for linear systems: {Theory} -- {Implementation} -- {Applications}},
  author={P. Van Overschee and B. De Moor},
  year={1996},
  publisher={Kluwer Academic Publishers}
}

@ARTICLE{koyama20,
	AUTHOR = {S. Koyama and G. Chardon and L. Daudet},
	TITLE = {Optimizing source and sensor placement for sound field control: an overview},
	JOURNAL = TransASLP_ACM,
	VOLUME = {28},
	PAGES = {696-714},
	YEAR = {2020}
}

@ARTICLE{kujawski24,
	AUTHOR = {A. Kujawski and A. J. R. Pelling and E. Sarradj},
	TITLE = {{MIRACLE}—a microphone array impulse response dataset for acoustic learning},
	JOURNAL = JASMP,
	VOLUME = {2024},
	NOTE = {Article No. 32},
	YEAR = {2024}
}

@INPROCEEDINGS{shen20,
	AUTHOR = {J. Shen and R. Duraiswami},
	TITLE = {Data-driven feedback delay network construction for real-time virtual room acoustics},
	BOOKTITLE = AM20,
	ADDRESS = {Graz, Austria},
	PAGES = {46-52},
	YEAR = {2020}
}

@INPROCEEDINGS{lyster22,
	AUTHOR = {S. V. Lyster and C. Erkut},
	TITLE = {A differentiable neural network approach to parameter estimation of reverberation},
	BOOKTITLE = SMC22,
	ADDRESS = {Saint-Étienne, France},
	PAGES = {358-364},
	YEAR = {2022}
}

@INPROCEEDINGS{bona22,
	AUTHOR = {R. Bona and D. Fantini and G. Presti and M. Tiraboschi and I. Engel and F. Avanzini},
	TITLE = {Automatic parameters tuning of late reverberation algorithms for audio augmented reality},
	BOOKTITLE = AM22,
	ADDRESS = {St. Pölten, Austria},
	PAGES = {36-43}, 
	YEAR = {2022}
}

@INPROCEEDINGS{ibnyahya22,
	AUTHOR = {I. Ibnyahya and J. D. Reiss},
	TITLE = {A method for matching room impulse responses with feedback delay networks},
	BOOKTITLE = AES153Conv, 
	ADDRESS = {New York, NY, USA},
	NOTE = {Express Paper No. 35},
	YEAR = {2022}
}

@article{sprunck2025,
	author = {T. Sprunck and A. Deleforge and Y. Privat and C. Foy},
 	title = {Fully reversing the shoebox image source method: From impulse responses to room parameters},
 	journal = TransASLP,
	pages = {1023-1033},
	year={2025}
}

@mastersthesis{deleforge2025,
    title={Can one hear the walls of a room? {Physics-} and data-driven inverse methods for acoustic signal processing},
    author={A. Deleforge},
    type = {{HDR} {Thesis}},
    year={2025},
    school={University of Strasbourg},
    address = {Strasbourg, France}
}

%

\end{document}